\documentclass[11pt]{article}
\usepackage[utf8]{inputenc}
\usepackage[margin=.8in]{geometry}
\usepackage{graphicx}
\usepackage{amsmath,lettrine}
\usepackage{gastex}

\newtheorem{definicao}{Definition}
\newtheorem{lema}{Lemma}      
\newtheorem{fato}{Fact}      

\newcommand{\tr}{\mbox{\rm\emph{trie}}}

\newcommand{\rk}[1]{\mbox{\rm{rank}}(#1)}
\newcommand{\sk}[1]{\mbox{\rm{sketch}}(#1)}

\newcommand{\trs}[1]{\mbox{\rm{BuscaTrie}}(#1)}
\newcommand{\dlt}[2]{\Delta(#1,#2)}

\begin{document}

\title{Fusion Tree Sorting}

\author{Luis A. A. Meira \\ \small School of Technology  \\ \small University of Campinas, Brazil \and Rogério H. B. de Lima \\ \small Institute of Science and Technology \\ \small Federal University of S\~{a}o  Paulo, Brazil }

\maketitle


\begin{abstract}
 The sorting problem is one of the most relevant problems in computer science. Within the scope of modern computer science it has been studied for more than 70 years. In spite of these facts, new sorting algorithms have been developed in recent years. Among several types of sorting algorithms, some are quicker; others are more economic in relation to space, whereas others insert a few restrictions in relation to data input. This paper is aimed at explaining the fusion tree data structure, which is responsible for the first sorting algorithm with complexity time smaller than $n\lg n$. The $n\lg n$ time complexity has led to some confusion and generated the wrong belief in part of the community of being the minimum possible for this type of problem.
\end{abstract}

\section{Introduction}

The sorting problem is perhaps the most studied problem in Computer Science. Its use is implicit in intermediate stages of almost all existing programs, such as database, spreadsheets, multimedia, etc. In addition, sorting has been studied by computer science for over 70 years. The currently and broadly used merge sort algorithm was proposed by Von Neumann in 1945~\cite{cormen}.
 
 The sorting problem consists of receiving a sequence $A= (a_1, \ldots , a_n)$ of $n$ numbers as input. The solution consists of a nondecreasing permutation  $A'= (a'_1, \ldots ,a'_n)$ of $A$. Although this work is focused on integers, the extension for rationals, floating point and character strings tend to be straight.
 

 All the sorting algorithms present characteristics that make them somehow more or less competitive in relation to their peers. Some of these characteristics are the sorting type, either stable or non-stable, extra space utilization for algorithm execution, and sorting time. Some algorithms can be quicker than others, depending on the characteristics of input data. For instance, selection sort tends to be advantageous when $n$ is small. Insertion sort tends to be rapid when the vector is partially sorted. Counting sort is advantageous when the difference between the maximum and the minimum element is limited.


The most broadly known sorting algorithms are comparison-based ones, such as merge sort, heap sort, insertion sort and quick sort, in addition to the counting-based ones, such as, for example, counting sort, bucket sort and radix sort. The counting-based algorithms require an input sequence with some restrictions. When such restrictions are satisfied, these algorithms can solve the sorting problem in linear time.

%
%

There is a lower bound of $\Omega(n\lg n) $ comparisons for sorting algorithms~\cite{knuth}. Such limit is based on a decision tree with $n!$ leaves, each of them representing an input vector permutation. Each permutation is a candidate to solve the problem. Provided that a comparison can distinguish two branches of a tree, a minimum of $\lg(n!) = \Theta(n\lg n) $ comparisons are required to sort a vector through a comparison-based sorting algorithm in the worst case. This lower bound was misinterpreted, thus generating a false belief in a part of the community in terms that sorting is a  $\Omega(n\lg n)$ problem. Such limit does not apply, for example, to algorithms using other operations rather than comparisons during the sorting process. The counting sort is able to sort a vector without performing any kind of comparison between the elements.


The algorithm under analysis in this paper is a comparison-based one and makes $\Theta(n \lg n)$  comparisons. However, $(\lg n)^{1/5}$ numbers are compared in $O(1)$. This means that multiple operations are performed in constant time. The following paragraph was extracted from~\cite{cormen}:


\begin{quote}
``The case of sorting $n$ $w$-bit integers in $o(n\lg n)$ time has been considered by many researchers. Several positive results have been obtained, each under slightly different assumptions about the model of computation and the restrictions placed on the algorithm. All the results assume that the computer memory is divided into addressable $w$-bit words. Fredman and Willard~\cite{fredman} introduced the fusion tree data structure and used it to sort $n$ integers in $O(n\lg n/\lg\lg n)$. This bound was later improved to $O(n\sqrt{\lg n})$ time by Andersson~\cite{arne1}. These algorithms require the use of multiplication and several precomputed constants. Andersson, Hagerup, Nilsson, and Raman~\cite{arne2} have shown how to sort $n$ integers in $O(n\lg\lg n) $ time without using multiplication, but their method requires storage that can be unbounded in terms of $n$. Using multiplicative hashing, we can reduce the storage needed to $O(n)$, but then the $O(n\lg\lg n)$  worst-case bound on the running time becomes an expected-time bound. Generalizing the exponential search trees of Andersson~\cite{arne1}, Thorup~\cite{thorup} gave an $O(n(\lg\lg n)^2)$-time sorting algorithm that does not use multiplication or randomization, and it uses linear space. Combining these techniques with some new ideas, 
Han~\cite{han} improved the bound for sorting to $O(n\lg\lg n\lg\lg\lg n)$ time. Although these algorithms are important theoretical breakthroughs, they are all fairly complicated and at the present time seem unlikely to compete with existing sorting algorithms in practice".
\end{quote}

Results: The sorting algorithm $O(n\lg n/\lg\lg n)$ under analysis in this paper is known to the literature. Our contribution consists of detailing the fusion tree data structure and the related sorting algorithm $O(n\lg n/\lg\lg n)$ proposed 
by~\cite{fredman}.


\subsection{Computational Model}

Consider a computer working with $w$-bit words. This computer is able to perform elementary operations such as addition, subtraction, multiplication, division, and remainders with $w$-bit integers in constant time. For example, a 64-bit computer has the capacity of processing 64 bits in constant time.


The general sorting case deals with integers with an arbitrary precision. For an integer with $mw$-bits, it is required $m$ accesses to the memory before completing the number reading. This paper works with the restricted sorting case where numbers are integers with $w$ bits. Such numbers are in the range $\{-2^{w-1},\ldots,2^{w-1}\}$  stored as binary integers, with 1 bit for the signal. Some special attention is needed to deal with repeated numbers. Thus, no repetition is assumed to simplify the explanation.


This work considered a computational model that is able to read and write any memory position in constant time, which is known as RAM memory. The RAM memory model is acceptable, though coexisting with the sequential access memory model. In the sequential access memory model, the tape needs to be moved up to the desired position before reading, thus spending linear time to read an integer. The merge sort algorithm is famous for keeping the complexity $O(n\lg n)$ even in a sequential memory model.

%

It is reasonable to assume that the computer is capable of processing $w=\log n$ bits in constant time. In the case of $n$ integers of $w$ bits in the memory, the maximum memory address will have at least $\lg n$ bits. The assumption that it is possible to access any position in the memory in constant time is equivalent to the computer processing 
addresses of $\lg n$ bits in constant time. Notice that the number of bits in the problem is $nw \geq n \lg n$. This means that one operation for each bit is $\Omega(n\lg n)$.


For a better understanding of the sorting process, we shall first show how to sort $n$ numbers using the B-tree data structure. The fusion tree data structure was proposed by~\cite{fredman} and it is a modified B-tree.


\section{B-Trees}

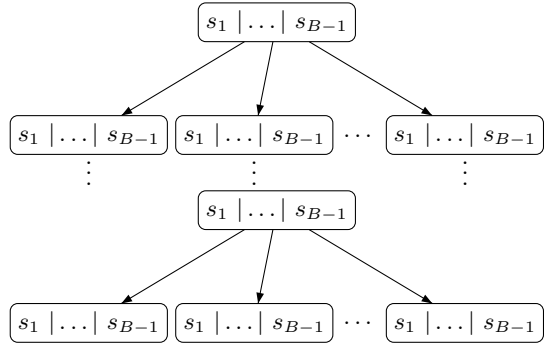
\begin{figure}[htb]
	\begin{center}
	\begin{picture}(50,50)(0,0)
	\node[Nadjust=wh,linecolor=white](CR)(0,26){\footnotesize $\vdots$} 
	\node[Nadjust=wh,linecolor=white](CR)(22,26){\footnotesize $\vdots$} 
	\node[Nadjust=wh,linecolor=white](CR)(50,26){\footnotesize $\vdots$} 	
	\node[Nadjust=wh,Nmr=1](A)(25,45){\footnotesize$s_1~|\ldots|~s_{B-1}$} 
	\node[Nadjust=wh,Nmr=1](B1)(0,30){\footnotesize $s_1~|\ldots|~s_{B-1}$} 
	\node[Nadjust=wh,Nmr=1](B2)(22,30){\footnotesize $s_1~|\ldots|~s_{B-1}$} 
	\node[Nadjust=wh,Nmr=1](B3)(50,30){\footnotesize $s_1~|\ldots|~s_{B-1}$} 
	\node[Nadjust=wh,linecolor=white](BR)(36,30){\footnotesize $\dots$} 
        \node[Nadjust=wh,linecolor=white](CR)(36,5){\footnotesize $\dots$} 
	\drawedge[ATnb=0,AHnb=1](A,B1){} 
	\drawedge[ATnb=0,AHnb=1](A,B2){} 
	\drawedge[ATnb=0,AHnb=1](A,B3){} 
	\node[Nadjust=wh,Nmr=1](D)(25,20){\footnotesize$s_1~|\ldots|~s_{B-1}$} 
	\node[Nadjust=wh,Nmr=1](C1)(0,5){\footnotesize $s_1~|\ldots|~s_{B-1}$} 
	\node[Nadjust=wh,Nmr=1](C2)(22,5){\footnotesize $s_1~|\ldots|~s_{B-1}$} 
	\node[Nadjust=wh,Nmr=1](C3)(50,5){\footnotesize $s_1~|\ldots|~s_{B-1}$} 
	\drawedge[ATnb=0,AHnb=1](D,C1){} 
	\drawedge[ATnb=0,AHnb=1](D,C2){} 
	\drawedge[ATnb=0,AHnb=1](D,C3){} 	 
	\end{picture}	
	\caption{A full B-tree data structure.}
	\label{fig_arvoreB}
	\end{center}
\end{figure}

B-trees are balanced search trees with a degree $t$, where $\frac{B}{2}~\leq~t~\leq~B$, for constant $B$. Each node contains a minimum of $\frac{B}{2}$ children and a maximum of $B$ children, except for the leaves, which contain no child, and the root-node, which does not present restriction in the minimum number of children. Each node has $t-1$ keys, and all the leaves are found in the same level. Notice that a degree-4 node has three keys. Figure~\ref{fig_arvoreB} illustrates a full B-tree, where all the nodes have $B - 1$ keys.


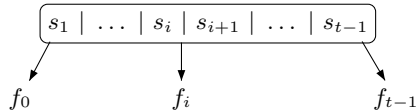
\begin{figure}[htb]
	\begin{center}
		\begin{picture}(50,15)(0,0)
	\node[Nadjust=wh,Nmr=1,linecolor=white](b1)(4.5,8.5){} 
		\node[Nadjust=wh,Nmr=1,linecolor=white](b2)(21.5,8.5){} 
			\node[Nadjust=wh,Nmr=1,linecolor=white](b3)(45,8.5){} 	
	\node[Nadjust=wh,Nmr=1](A)(25,10){\footnotesize$s_1~|~\ldots~|~s_i~|~s_{i+1}~|~\ldots~|~s_{t-1}$} 
	\node[Nadjust=wh,Nmr=1,linecolor=white](f1)(0,0){\footnotesize$f_0$} 
	\node[Nadjust=wh,Nmr=1,linecolor=white](f2)(21.5,0){\footnotesize$f_i$} 
	\node[Nadjust=wh,Nmr=1,linecolor=white](f3)(50,0){\footnotesize$f_{t-1}$} 
	\drawedge[ATnb=0,AHnb=1](b1,f1){} 
	\drawedge[ATnb=0,AHnb=1](b2,f2){} 
	\drawedge[ATnb=0,AHnb=1](b3,f3){} 	
	\end{picture}	\caption{The structure of a B-tree node.}
	\label{fig_noArvoreB}
	\end{center}
\end{figure}

In addition, the B-tree respects the following property: Each non-root node has $t-1$ sorted elements 
$S=(s_1, \ldots, s_{t-1})$. Each non-leaf and non-root node has $t$ children  ($f_0,\ldots, f_{t-1})$ where each child is a B-tree. The elements in the $f_0$ tree are smaller than $s_1$. The elements in $f_i$ are greater
 than $s_i$ and smaller than $s_{i+1}$. The elements in $f_{t-1}$ are all greater than $s_{t-1}$. See Figure~\ref{fig_noArvoreB}. 


Searching a key $k \notin S$ in a B-tree node requires finding the correct child $X$ to continue the search. If $k < s_1$, the search continues int $f_0$ child. If $k > s_{t-1}$, the search continues in $f_{t-1}$ child. If $s_i < k < s_{i+1}$, the search continues in $f_i$ child, between $s_i$ and $s_{i+1}$. The B-tree operations complexity time are related to its height. The following lemma is based in~\cite{cormen}.

%

\begin{lema}
A B-tree with degree $B \geq 4$  and height $h$ respect:

\[ h =O (\log_B n) \]

\end{lema}

A sequential search is made to search a key $k$ in a B-tree node. Such search takes $O(B)$ and it is repeated in each B-tree level in the worst case. The result is an $O(B \log_B {n})$ overall time to search the key. As $B$ is constant, the complexity is equivalent to $O (\lg n)$.

\begin{figure}[htb]
	\begin{center}
\begin{picture}(81,25)(0,0)
	\node[Nframe=n,Nadjust=wh,Nmr=1](i6)(6,9){} 
	\node[Nframe=n,Nadjust=wh,Nmr=1](f6)(6,1){\tiny$T_1$} 
	\drawedge[ATnb=0,AHnb=1](i6,f6){} 
	\node[Nframe=n,Nadjust=wh,Nmr=1](i1)(17,19.1){} 
	\node[Nframe=n,Nadjust=wh,Nmr=1](f1)(17,11.2){} 
	\drawedge[ATnb=0,AHnb=1](i1,f1){} 
	\node[Nframe=n,Nadjust=wh,Nmr=1](i20)(63.5,18.8){} 
	\node[Nframe=n,Nadjust=wh,Nmr=1](f20)(58,11.2){} 
	\drawedge[ATnb=0,AHnb=1](i20,f20){} 
	\node[Nframe=n,Nadjust=wh,Nmr=1](i21)(65.5,18.8){} 
	\node[Nframe=n,Nadjust=wh,Nmr=1](f21)(72,11.1){} 
	\drawedge[ATnb=0,AHnb=1](i21,f21){} 	
	\node[Nframe=n,Nadjust=wh,Nmr=1](i7)(21.5,8.7){} 
	\node[Nframe=n,Nadjust=wh,Nmr=1](f7)(21.5,1){\tiny$T_6$} 
	\drawedge[ATnb=0,AHnb=1](i7,f7){} 
	\node[Nframe=n,Nadjust=wh,Nmr=1](i8)(25,8.7){} 
	\node[Nframe=n,Nadjust=wh,Nmr=1](f8)(25,1){\tiny$T_7$} 
	\drawedge[ATnb=0,AHnb=1](i8,f8){} 
	\node[Nframe=n,Nadjust=wh,Nmr=1](i9)(28,9){} 
	\node[Nframe=n,Nadjust=wh,Nmr=1](f9)(28,1){\tiny$T_8$} 
	\drawedge[ATnb=0,AHnb=1](i9,f9){} 
	\node[Nframe=n,Nadjust=wh,Nmr=1](i2)(18.5,8.7){} 
	\node[Nframe=n,Nadjust=wh,Nmr=1](f2)(18.5,1){\tiny$T_5$} 
	\drawedge[ATnb=0,AHnb=1](i2,f2){} 
	\node[Nframe=n,Nadjust=wh,Nmr=1](i3)(15.5,8.7){} 
	\node[Nframe=n,Nadjust=wh,Nmr=1](f3)(15.5,1){\tiny$T_4$} 
	\drawedge[ATnb=0,AHnb=1](i3,f3){} 
	\node[Nframe=n,Nadjust=wh,Nmr=1](i4)(12.5,8.7){} 
	\node[Nframe=n,Nadjust=wh,Nmr=1](f4)(12.5,1){\tiny$T_3$} 
	\drawedge[ATnb=0,AHnb=1](i4,f4){} 
	\node[Nframe=n,Nadjust=wh,Nmr=1](i5)(9,8.7){} 
	\node[Nframe=n,Nadjust=wh,Nmr=1](f5)(9,1){\tiny$T_2$} 
		\drawedge[ATnb=0,AHnb=1](i5,f5){} 
	\node[Nframe=n,Nadjust=wh,Nmr=1](i17)(53.5,8.7){} 
	\node[Nframe=n,Nadjust=wh,Nmr=1](f17)(53.5,1){\tiny$T_1$} 
	\drawedge[ATnb=0,AHnb=1](i17,f17){} 
	\node[Nframe=n,Nadjust=wh,Nmr=1](i18)(56.5,8.7){} 
	\node[Nframe=n,Nadjust=wh,Nmr=1](f18)(56.5,1){\tiny$T_2$} 
	\drawedge[ATnb=0,AHnb=1](i18,f18){} 
	\node[Nframe=n,Nadjust=wh,Nmr=1](i19)(60,8.7){} 
	\node[Nframe=n,Nadjust=wh,Nmr=1](f19)(60,1){\tiny$T_3$} 
	\drawedge[ATnb=0,AHnb=1](i19,f19){} 
	\node[Nframe=n,Nadjust=wh,Nmr=1](i12)(63,9){} 
	\node[Nframe=n,Nadjust=wh,Nmr=1](f12)(63,1){\tiny$T_4$} 
	\drawedge[ATnb=0,AHnb=1](i12,f12){} 
	\node[Nframe=n,Nadjust=wh,Nmr=1](i13)(67,9.2){} 
	\node[Nframe=n,Nadjust=wh,Nmr=1](f13)(67,1){\tiny$T_5$} 
	\drawedge[ATnb=0,AHnb=1](i13,f13){} 
	\node[Nframe=n,Nadjust=wh,Nmr=1](i14)(70.4,9){} 
	\node[Nframe=n,Nadjust=wh,Nmr=1](f14)(70.4,1){\tiny$T_6$} 
	\drawedge[ATnb=0,AHnb=1](i14,f14){} 
	\node[Nframe=n,Nadjust=wh,Nmr=1](i15)(74,9){} 
	\node[Nframe=n,Nadjust=wh,Nmr=1](f15)(74,1){\tiny$T_7$} 	
	\drawedge[ATnb=0,AHnb=1](i15,f15){} 
	\node[Nframe=n,Nadjust=wh,Nmr=1](i16)(77,9.3){} 
	\node[Nframe=n,Nadjust=wh,Nmr=1](f16)(77,1){\tiny$T_8$} 	
	\drawedge[ATnb=0,AHnb=1](i16,f16){} 		
		\node[Nframe=n,Nadjust=wh](x)(30,15){} 	
		\node[Nframe=n,Nadjust=wh](y)(50,15){} 	
		\drawedge[ATnb=0,AHnb=1,linewidth=.4,AHLength=2.5](x,y){Split} 	
		\drawedge[ELside=r](x,y){$O(B)$} 			
	\node[Nadjust=wh,Nmr=1](A1)(17,20){\footnotesize $\ldots$ N W $\ldots$} 
	\node[Nadjust=wh,Nmr=1](B1)(17,10){\footnotesize P Q R {\bf S} T U V} 
	\node[Nadjust=wh,Nmr=1](A2)(65,20){\footnotesize $\ldots$ N {\bf S} W $\ldots$} 
	\node[Nadjust=wh,Nmr=1](A)(58,10){\footnotesize P Q R} 
	\node[Nadjust=wh,Nmr=1](A)(72,10){\footnotesize T U V} 
	\end{picture}	

	\caption{Key insertion int a complete B-tree node~\cite{cormen}.}
	\label{fig_separaArvoreB}
	\end{center}
\end{figure}
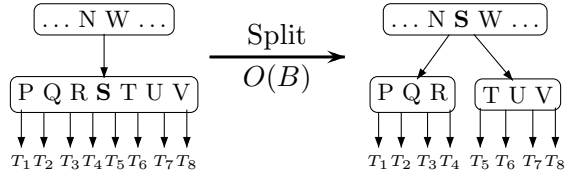

The key insertion needs an initial search to find the recipient node. If such node is incomplete, the key can be accommodated into the node in $O(B)$. It is the cost to insert an element in a central position of a vector with $B$ elements. If the recipient node is full, it must be split. See Figure~\ref{fig_separaArvoreB}.  Let $s_m$ be the vector median. Such element is inserted in the parent node. One node is created with the elements smaller than $s_m$ and other with the elements greater than $s_m$. Such nodes become the left and the right child of $s_m$ respectively. Both nodes have exactly $\frac{B}{2}-1$ keys.


A vector can be split in half in $O(B)$ through elementary operations. If the parent node is complete, it must be also split. Such process can propagate up to the root.


If such process propagate to root, the split will have complexity $O(B h) = O(B  \log_B n)$. Such worst case complexity can be improved by an amortized analysis. Each split has an $O(B)$ cost and it creates an additional node. As the insertion of $n$ elements creates $O(\frac{n}{B})$ nodes, the overall cost will be $O(n)$. The split cost considering the insertion of $n$ element takes $O(n)$, although a single insertion can waste $O(\lg n)$.
Ignoring the split cost, each insertion will cost $O(B\log_B n)$, which is the cost to search a key in the tree plus the cost to insert a key in a node.

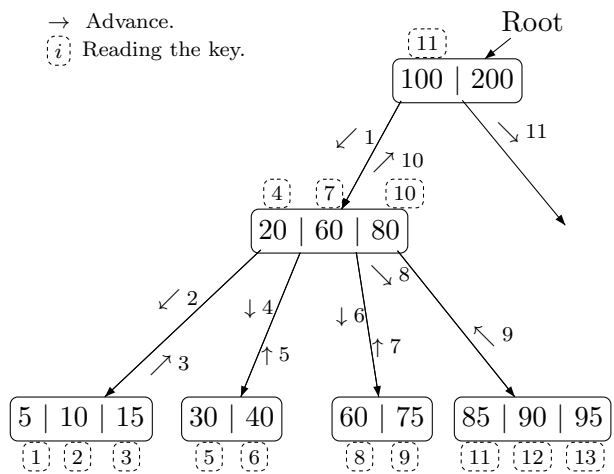
\begin{figure}[h!]
	\begin{center}
	\begin{picture}(80,60)(0,0)
	\node[Nframe=n,Nadjust=wh,Nmr=1](R)(70,58){Root} 
	\node[Nadjust=wh,Nmr=1](n1)(60,50){$100~|~200$} 
	\node[Nframe=n,Nadjust=wh,Nmr=1](i1)(34.6,28.2){} 
	\node[Nframe=n,Nadjust=wh,Nmr=1](i2)(39.5,28.1){} 
	\node[Nframe=n,Nadjust=wh,Nmr=1](i3)(46.4,28.2){} 
	\node[Nframe=n,Nadjust=wh,Nmr=1](i4)(51.4,28.2){} 
	\node[Nframe=n,Nadjust=wh,Nmr=1](i5)(53,48.2){} 	
	\node[Nframe=n,Nadjust=wh,Nmr=1](i6)(60,48.2){} 	
	\node[dash={.5}0,Nadjust=wh,Nmr=1](x1)(7,54){\scriptsize$i$} 
	\node[dash={.5}0,Nadjust=wh,Nmr=1](x1)(4,0){\scriptsize$1$} 
	\node[dash={.5}0,Nadjust=wh,Nmr=1](x1)(9.5,0){\scriptsize$2$} 
	\node[dash={.5}0,Nadjust=wh,Nmr=1](x1)(16,0){\scriptsize$3$} 	
	\node[dash={.5}0,Nadjust=wh,Nmr=1](x1)(27,0){\scriptsize$5$} 	
	\node[dash={.5}0,Nadjust=wh,Nmr=1](x1)(33,0){\scriptsize$6$} 	
	\node[dash={.5}0,Nadjust=wh,Nmr=1](x1)(47,0){\scriptsize$8$} 	
	\node[dash={.5}0,Nadjust=wh,Nmr=1](x1)(53,0){\scriptsize$9$} 			
	\node[dash={.5}0,Nadjust=wh,Nmr=1](x1)(63,0){\scriptsize$11$}
	\node[dash={.5}0,Nadjust=wh,Nmr=1](x1)(70,0){\scriptsize$12$}
	\node[dash={.5}0,Nadjust=wh,Nmr=1](x1)(77,0){\scriptsize$13$}
	\node[dash={.5}0,Nadjust=wh,Nmr=1](x1)(36,35){\scriptsize$4$} 	
	\node[dash={.5}0,Nadjust=wh,Nmr=1](x1)(43,35){\scriptsize$7$} 	
	\node[dash={.5}0,Nadjust=wh,Nmr=1](x1)(53,35){\scriptsize$10$} 		
	\node[dash={.5}0,Nadjust=wh,Nmr=1](x1)(56,55){\scriptsize$11$} 
			
	\node[Nframe=n,Nadjust=wh,Nmr=1](X)(20,58){\scriptsize$\rightarrow$ ~Advance.~~~~~~~~~~~~} 	
	\node[Nframe=n,Nadjust=wh,Nmr=1](X)(20,54){\scriptsize ~~~~~~ Reading the key.~~~~~} 
	\drawedge[ATnb=0,AHnb=1](R,n1){} 
	\node[Nadjust=wh,Nmr=1](n2)(43,30){$20~|~60~|~80$} 
	\node[Nframe=n,Nadjust=wh,Nmr=1](n3)(75,30){} 
	\drawedge[ATnb=0,AHnb=1,ELpos=40,ELside=r,ELdist=0](i5,n2){\scriptsize$\swarrow 1$} 
	\drawedge[ATnb=0,AHnb=1,ELpos=40,ELdist=0](i5,n2){\scriptsize$\nearrow 10$} 
	\drawedge[ATnb=0,AHnb=1,ELpos=40,ELdist=0](i6,n3){\scriptsize$\searrow 11$} 
	\node[Nadjust=wh,Nmr=1](n4)(10,5){$5~|~10~|~15$} 
	\node[Nadjust=wh,Nmr=1](n6)(30,5){$30~|~40$} 
	\node[Nadjust=wh,Nmr=1](n7)(50,5){$60~|~75$} 
	\node[Nadjust=wh,Nmr=1](n9)(70,5){$85~|~90~|~95$} 
	
	\drawedge[ATnb=0,AHnb=1,ELpos=40,ELside=r,ELdist=0](i1,n4){\scriptsize$\swarrow 2$}
	\drawedge[ATnb=0,AHnb=0,ELpos=40,ELside=r,ELdist=0](n4,i1){\scriptsize$\nearrow 3$}

	\drawedge[ATnb=0,AHnb=1,ELpos=40,ELside=r,ELdist=0](i2,n6){\scriptsize$\downarrow 4$}
	\drawedge[ATnb=0,AHnb=0,ELpos=40,ELside=r,ELdist=0](n6,i2){\scriptsize$\uparrow 5$}

	\drawedge[ATnb=0,AHnb=1,ELpos=40,ELside=r,ELdist=0](i3,n7){\scriptsize$\downarrow 6$}
	\drawedge[ATnb=0,AHnb=0,ELpos=40,ELside=r,ELdist=0](n7,i3){\scriptsize$\uparrow 7$}

	\drawedge[ATnb=0,AHnb=1,ELpos=10,ELside=r,ELdist=0](i4,n9){\scriptsize$\searrow 8$}
	\drawedge[ATnb=0,AHnb=0,ELpos=40,ELside=r,ELdist=0](n9,i4){\scriptsize$\nwarrow 9$}

	\end{picture}	
	\caption{B-tree in-order traversal.}
	\label{fig_pesquisaOrdenadaArvoreB}
	\end{center}
\end{figure}

To sort a sequence with $n$ elements using a B-tree, all elements must be inserted in an initially empty tree.
An in-order traversal result in a sorted sequence. See Figure~\ref{fig_pesquisaOrdenadaArvoreB}. The solid arrows represent the tree traversal path. The dashed arrows represent the reading sequence. Each arrow has an integer representing the traversal sequence.


The complexity time to sort $n$ integers is the sum of the time to insert $n$ keys in the tree that is $O(n  B \log_B n)$. If $B$ is a constant, such complexity will be $O(n \lg n)$.


\section{Fusion Tree}

This section describes the fusion tree data structure proposed by~\cite{fredman}. A fusion tree is similar to a B-tree in many aspects. One difference between B-tree and the fusion tree is the B value. In a B-tree the $B$ value is a constant while in a fusion tree the $B$ is a function of $n$. More precisely,  $B=(\lg n)^\frac{1}{5}$. Another difference is the time to search a key in a node. The B-tree uses O(B) operations while the fusion tree uses O(1) operations to search a key $k$ in a node.


Consider the problem of finding the predecessor or the successor of a key $x$ in a set $S$. Such problem consists in finding the number immediately above or below $x$ in $S$. Fusion tree is a data structure similar to B-tree but it solves the predecessor and successor problem in O(1) inside a node. Given a search key $x$, the fusion tree is able to find the child branch relative to $x$ in constant time despite the fact that the size of $S$ increases with $n$.


The following notation present in ~\cite{fredman} is needed:


\begin{definicao}{$\rk{x}$:}
Given a set of integer numbers $S$ and an integer $x$, let $\rk{x}$ be the value $|\{t~ |~ t\in S, t \leq x\}|$. In other 
words, $\rk{x}$ represents the number of elements smaller than or equal $x$.
\end{definicao}

The problem of sort $n$ number is equivalent to finding $\rk{x}$ for all $x$. Such function provides the exact $x$ position in the sorted vector. Moreover, $\rk{x}$ provides the correct child to continue the search of an element $x$ in a B-tree node. Fusion tree is based in a trie data structure described in~\cite{ajtai}. Next subsection is devoted to the trie data structure.


\subsection{Trie data structure~\cite{ajtai}}

Let a trie be a binary tree with the following construction rule. Given a $w$-bit integer $x$, each bit of $x$ is a node in the trie. If the most significant bit of $x$ is $0$, $x$ is a left root child. If it is $1$, $x$ is a right root child. Such property is recursively applied to each bit of $x$.

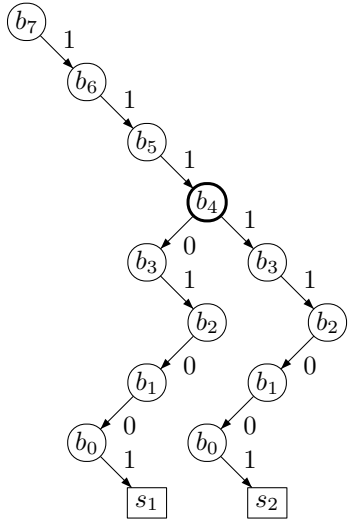
\begin{figure}[htb]
	\begin{center}
	\setlength{\unitlength}{.8mm}
	\begin{picture}(70,90)(0,0)
	\node[Nadjust=wh,Nmr=3](b0)(10,85){\small $b_7$} 
	\node[Nadjust=wh,Nmr=3](b1)(20,75){\small $b_6$} 
	\node[Nadjust=wh,Nmr=3](b2)(30,65){\small $b_5$} 
	\node[Nadjust=wh,Nmr=3,linewidth=0.5](b3)(40,55){\small $b_4$} 
	\node[Nadjust=wh,Nmr=3](b4)(50,45){\small $b_3$} 
	\node[Nadjust=wh,Nmr=3](c4)(30,45){\small $b_3$} 
	\node[Nadjust=wh,Nmr=3](b5)(60,35){\small $b_2$} 
	\node[Nadjust=wh,Nmr=3](c5)(40,35){\small $b_2$} 
	\node[Nadjust=wh,Nmr=3](b6)(50,25){\small $b_1$} 
	\node[Nadjust=wh,Nmr=3](c6)(30,25){\small $b_1$} 
	\node[Nadjust=wh,Nmr=3](b7)(40,15){\small $b_0$} 
	\node[Nadjust=wh,Nmr=3](c7)(20,15){\small $b_0$} 
	\node[Nadjust=wh,Nmr=0](c8)(30,5){\small $s_1$} 
	\node[Nadjust=wh,Nmr=0](b8)(50,5){\small $s_2$} 
	
	\drawedge[ATnb=0,AHnb=1](b0,b1){\small 1} 
	\drawedge[ATnb=0,AHnb=1](b1,b2){\small 1}  
	\drawedge[ATnb=0,AHnb=1](b2,b3){\small 1}
	\drawedge[ATnb=0,AHnb=1](b3,b4){\small 1}
	\drawedge[ATnb=0,AHnb=1](b4,b5){\small 1} 
	\drawedge[ATnb=0,AHnb=1](b5,b6){\small 0}   
	\drawedge[ATnb=0,AHnb=1](b6,b7){\small 0} 
	\drawedge[ATnb=0,AHnb=1](b7,b8){\small 1} 
	\drawedge[ATnb=0,AHnb=1](b3,c4){\small 0}
	\drawedge[ATnb=0,AHnb=1](c4,c5){\small 1} 
	\drawedge[ATnb=0,AHnb=1](c5,c6){\small 0}   
	\drawedge[ATnb=0,AHnb=1](c6,c7){\small 0} 
	\drawedge[ATnb=0,AHnb=1](c7,c8){\small 1} 			
	\end{picture}	
	\setlength{\unitlength}{1mm}
	\caption{
	Trie data structure for $s_1$ and $s_2$. It is enough to compare $b_4$ to sort $s_1$ and $s_2$.
	}
	\label{fig_trieBit}
	\end{center}
\end{figure}

Given an arbitrary integer $i$, let $b_i$ be the $i$-th least significant bit. Thus, $b_0$ is the least significant bit, $b_1$ is the second least significant bit and so on. Consider two binary integers $s_1 = 11101001$ and $s_2 = 11111001$. 
Figure~\ref{fig_trieBit} shows a trie with $s_1$ and $s_2$. The trie leaves are always sorted. We consider $\rk{x}$ calculated for all element in the trie.


Consider the following definition:
\begin{definicao}{$\dlt{s_1}{s_2}$:}
Given two integers  $s_1$ and $s_2$, let $\dlt{s_1}{s_2}$ be the relevant bit between $s_1$ and $s_2$, meaning the most significant bit that diverges between $s_1$ and $s_2$.
\end{definicao}

When a set of integers are compared, some bits are irrelevant and can be discarded. Only a few-bit, named relevant bits, are sufficient to sort a set of integers. 

\begin{definicao}{\textbf{relevant bits}:} Consider a trie with a set of elements S. The relevant bits of S are the bits for which there is a branch in the trie.
\end{definicao}

\begin{figure}[htb]
	\begin{center}
	\setlength{\unitlength}{.8mm}
	\begin{picture}(70,30)(0,0)
	\node[Nadjust=wh,Nmr=3](b3)(35,20){\small $b_4$} 
	\node[Nframe=n,Nadjust=wh,Nmr=3](x)(65,20){\small relevant bit} 
	
	\node[Nadjust=wh,Nmr=0](c8)(25,5){\small $s_1$} 
	\node[Nadjust=wh,Nmr=0](b8)(45,5){\small $s_2$} 
	\drawedge[ATnb=0,AHnb=1](b3,b8){\small 1}
	\drawedge[ATnb=0,AHnb=1,ELside=r](b3,c8){\small 0}
	\drawedge[ATnb=0,AHnb=1](x,b3){}
	\end{picture}	
	\setlength{\unitlength}{1mm}
	\caption{Compressed trie used to compare $s_1$ and $s_2$ with the relevant bit.}
	\label{fig_trieBitCondensado}
	\end{center}
\end{figure}
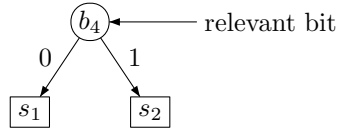

In the previous example, to compare and sort the binary number $s_1=11101001$ e $s_2=11111001$, it is sufficient to compare the most significant bit that diverges between $s_1$ and $s_2$. Considering the previous numbers, such bit is $b_4$, with values 0 in $s_1$ and $1$ in $s_2$. Thus, $\dlt{s_1}{s_2}=b_4$. Such bit is used to conclude that $s_1$ is greater than $s_2$. All other bits are irrelevant. See Figure~\ref{fig_trieBitCondensado}.


Let $\oplus$ be a bitwise XOR between two words. Given two integers $s_1$  and $s_1$ , $\dlt{s_1}{s_2}$ can be obtained as:
%
%
$$\dlt{s_1}{s_2}=\lfloor\lg (s_1 \oplus  s_2)\rfloor.$$

Consider an integer sequence $S = (s_1, \ldots , s_t)$ and a trie data structure. After inserting all $S$ elements in the trie, a compression will be performed where all irrelevant bits will be discarded. Such new tree will be named compressed trie.


The compressed trie has each element of $S$ as a leaf. The internal nodes store the respective relevant bit.


\begin{lema}
Given a compressed trie with $S=(s_1,\ldots,s_t)$, the numbers of relevant bit will be at most~$t-1$.
\label{nbits}
\end{lema}

Lemma~\ref{nbits}  is correct because each relevant bit is related to a branch in the compressed trie. The number of branches will be exactly $t-1$. Eventually, two distinct branches can occur at the same level.


To search a key $x$ in a compressed trie, each bit of the trie is compared with the correspondent $x$ bit from the root to the leaves. In each node, if the $x$ bit is zero, the search continues in the left branch. If the bit is 1, the search continues in the right branch. Figure~\ref{fig_pesquisaTrie}  illustrates a search of a key $x$ in a compressed trie with elements $a$, $b$, $c$ and $d$. Let $\trs{x}$ be such search result. In Figure~\ref{fig_pesquisaTrie},  $\trs{x}=c$.

%

Ao realizar esse procedimento, chega-se a uma folha, que é um elemento $s$ de $S$. Isso  mostra que $x$ e $s$ têm os mesmos bits nas posições percorridas no caminho da $\tr$.

\begin{figure*}[htb]
	\begin{center}
	\setlength{\unitlength}{.9mm}
\begin{picture}(120,80)(0,0)
	\put(70,65){\begin{tabular}{cc}
	  S=\{a,b,c,d\}~~~~\\
	  a = 1 1 {\bf 0 1} 1 1 1 {\bf 1}\\
	  b = 1 1 {\bf 1 0}  0 0 0 {\bf 0}\\
	  c = 1 1 {\bf 1 0} 0 0 0 {\bf 1}\\
	  d = 1 1 {\bf 1 1} 1 1 1 {\bf 0}\\
	 x = 1 1 {\bf 1 0} 0 1 1 {\bf 1}			
	\end{tabular}}
	\node[Nadjust=wh,Nmr=3](fb2)(85,45){\small $b_5$} 
	\node[Nadjust=wh,Nmr=3](fb3)(93,35){\small $b_4$} 
	\node[Nadjust=wh,Nmr=3,linewidth=0.5](fb5)(85,25){\small $b_2$} 
	\node[Nadjust=wh,Nmr=3](fb7)(77,15){\small $b_0$} 
	\node[Nadjust=wh,Nmr=0](fc)(99,25){\small d} 
	\node[Nadjust=wh,Nmr=0](fb)(71,5){\small b} 
	\node[Nadjust=wh,Nmr=0](fd)(85,5){\small c} 
	\node[Nadjust=wh,Nmr=0](x)(93,15){\small x} 
	\node[Nadjust=wh,Nmr=0](fa)(77,35){\small $a$} 
	\drawedge[ATnb=0,AHnb=1,ELside=r](fb3,fb5){\small $0$}
	\drawedge[ATnb=0,AHnb=1,ELside=r](fb5,fb7){\small $0$}
	\drawedge[ATnb=0,AHnb=1,ELside=l](fb3,fc){\small 1}
	\drawedge[ATnb=0,AHnb=1,ELside=l](fb5,x){\small 1}
	\drawedge[ATnb=0,AHnb=1,ELside=l](fb7,fd){\small $1$}
	\drawedge[ATnb=0,AHnb=1,ELside=r](fb7,fb){\small 0}
	\drawedge[ATnb=0,AHnb=1,ELside=l](fb2,fb3){\small $1$}
	\drawedge[ATnb=0,AHnb=1,ELside=r](fb2,fa){\small 0}	
	
		\node[Nadjust=wh,Nmr=3](b0)(10,75){\small $b_7$} 
	\node[Nadjust=wh,Nmr=3](b1)(20,70){\small $b_6$} 
	\node[Nadjust=wh,Nmr=3](b2)(30,65){\small $b_5$} 
	\node[Nadjust=wh,Nmr=3](b3)(40,55){\small $b_4$} 
	\node[Nadjust=wh,Nmr=3](a3)(00,55){\small $b_4$} 
	\node[Nadjust=wh,Nmr=3](a4)(3,45){\small $b_3$} 
	\node[Nadjust=wh,Nmr=3](a5)(6,35){\small $b_2$} 
	\node[Nadjust=wh,Nmr=3](a6)(9,25){\small $b_1$} 
	\node[Nadjust=wh,Nmr=3](a7)(12,15){\small $b_0$} 
	\node[Nadjust=wh,Nmr=0](a8)(15,5){\small a} 
	\drawedge[ATnb=0,AHnb=1,ELside=r](b2,a3){\small 0} 
	\drawedge[ATnb=0,AHnb=1](a3,a4){\small 1} 
	\drawedge[ATnb=0,AHnb=1](a4,a5){\small 1} 	
	\drawedge[ATnb=0,AHnb=1](a5,a6){\small 1} 	
	\drawedge[ATnb=0,AHnb=1](a6,a7){\small 1} 	
	\drawedge[ATnb=0,AHnb=1](a7,a8){\small 1} 
	
	\node[Nadjust=wh,Nmr=3](b4)(47,45){\small $b_3$} 
	\node[Nadjust=wh,Nmr=3](b5)(50,35){\small $b_2$} 
	\node[Nadjust=wh,Nmr=3](b6)(53,25){\small $b_1$} 
	\node[Nadjust=wh,Nmr=3](b7)(56,15){\small $b_0$} 
	\node[Nadjust=wh,Nmr=0](b8)(53,5){\small d} 
	
	\node[Nadjust=wh,Nmr=3](c4)(37,45){\small $b_3$}

	\node[Nadjust=wh,Nmr=3,linewidth=0.5](c5)(34,35){\small $b_2$} 
	\node[Nadjust=wh,Nmr=3](c6)(31,25){\small $b_1$} 
	\node[Nadjust=wh,Nmr=3](x6)(40,25){\small $b_1$} 
	\node[Nadjust=wh,Nmr=3](x7)(42,15){\small $b_0$} 
	\node[Nadjust=wh,Nmr=0](x8)(45,5){\small x} 
	\drawedge[ATnb=0,AHnb=1](c5,x6){\small 1} 
	\drawedge[ATnb=0,AHnb=1](x6,x7){\small 1} 
	\drawedge[ATnb=0,AHnb=1](x7,x8){\small 1} 			

	\node[Nadjust=wh,Nmr=3](c7)(28,15){\small $b_0$} 
	\node[Nadjust=wh,Nmr=0](c8)(25,5){\small b} 
	\node[Nadjust=wh,Nmr=0](c9)(31,5){\small c}

	\drawedge[ATnb=0,AHnb=1](b0,b1){\small 1} 
	\drawedge[ATnb=0,AHnb=1](b1,b2){\small 1}  
	\drawedge[ATnb=0,AHnb=1](b2,b3){\small 1}
	\drawedge[ATnb=0,AHnb=1](b3,b4){\small 1}
	\drawedge[ATnb=0,AHnb=1](b4,b5){\small 1} 
	\drawedge[ATnb=0,AHnb=1](b5,b6){\small 1}   
	\drawedge[ATnb=0,AHnb=1](b6,b7){\small 1} 
	\drawedge[ATnb=0,AHnb=1](b7,b8){\small 0} 
	\drawedge[ATnb=0,AHnb=1,ELside=r](b3,c4){\small 0}
	\drawedge[ATnb=0,AHnb=1,ELside=r](c4,c5){\small 0} 
	\drawedge[ATnb=0,AHnb=1,ELside=r](c5,c6){\small 0}   
	\drawedge[ATnb=0,AHnb=1,ELside=r](c6,c7){\small 0} 
	\drawedge[ATnb=0,AHnb=1,ELside=r](c7,c8){\small 0} 
	\drawedge[ATnb=0,AHnb=1,ELside=l](c7,c9){\small 1} 	
	
	\end{picture}	
	\setlength{\unitlength}{1mm}
	\caption{ Trie and compressed trie after the $x$ insertion~\cite{ajtai}.}
	\label{fig_insercaoTrie}
	\end{center}
\end{figure*}
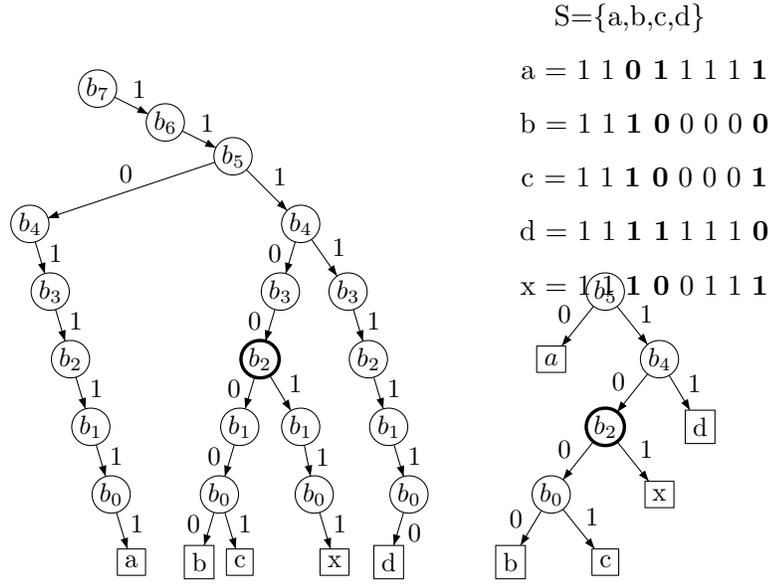

  No exemplo da Figura~\ref{fig_pesquisaTrie} nota-se que o primeiro bit a divergir entre $x$ e $c$ é o $b_2$,
  ou seja $\dlt{x}{c}=b_2$.
  A Figura~\ref{fig_insercaoTrie} mostra como a $\tr$ fica após a inserção do elemento $x$ na estrutura.

\begin{figure*}[htb]
	\begin{center}
\begin{picture}(120,80)(0,0)
	\put(70,60){\begin{tabular}{cc}
	  S=\{a,b,c,d\}~~~~\\
	  a = 1 1 {\bf 0 1} 1 1 1 {\bf 1}\\
	  b = 1 1 {\bf 1 0}  0 0 0 {\bf 0}\\
	  c = 1 1 {\bf 1 0} 0 0 0 {\bf 1}\\
	  d = 1 1 {\bf 1 1} 1 1 1 {\bf 0}\\
	~\\		
	 x = 1 1 {\bf 1 0} 0 1 1 {\bf 1}			
	\end{tabular}}
	\node[Nadjust=wh,Nmr=3](fb2)(85,35){\small $b_5$} 
	\node[Nadjust=wh,Nmr=3](fb3)(95,25){\small $b_4$} 
	\node[Nadjust=wh,Nmr=3](fb7)(85,15){\small $b_0$} 
	\node[Nadjust=wh,Nmr=0](fc)(105,15){\small $d$} 
	\node[Nadjust=wh,Nmr=0](fb)(75,5){\small $b$} 
	\node[Nadjust=wh,Nmr=0,linewidth=.4](fd)(95,5){\small $c$} 
	\node[Nframe=n,Nadjust=wh,Nmr=0](x)(100,5){\small $\mathbf{x}$} 
	\node[Nadjust=wh,Nmr=0](fa)(75,25){\small $a$} 
	\drawedge[ATnb=0,AHnb=1,ELside=r,linewidth=.5,AHLength=2,ELpos=70](fb3,fb7){\small $x[4]=0$}
	\drawedge[ATnb=0,AHnb=1,ELside=l](fb3,fc){\small 1}
	\drawedge[ATnb=0,AHnb=1,ELside=l,linewidth=.5,AHLength=2,ELpos=70](fb7,fd){\small $x[0]=1$}
	\drawedge[ATnb=0,AHnb=1,ELside=r](fb7,fb){\small 0}
	\drawedge[ATnb=0,AHnb=1,ELside=l,linewidth=.5,AHLength=2,ELpos=70](fb2,fb3){\small $x[5]=1$}
	\drawedge[ATnb=0,AHnb=1,ELside=r](fb2,fa){\small 0}	
	
		\node[Nadjust=wh,Nmr=3](b0)(10,75){\small $b_7$} 
	\node[Nadjust=wh,Nmr=3](b1)(20,70){\small $b_6$} 
	\node[Nadjust=wh,Nmr=3,linewidth=0.5](b2)(30,65){\small $b_5$} 
	\node[Nadjust=wh,Nmr=3,linewidth=0.5](b3)(40,55){\small $b_4$} 
	\node[Nadjust=wh,Nmr=3](a3)(00,55){\small $b_4$} 
	\node[Nadjust=wh,Nmr=3](a4)(3,45){\small $b_3$} 
	\node[Nadjust=wh,Nmr=3](a5)(6,35){\small $b_2$} 
	\node[Nadjust=wh,Nmr=3](a6)(9,25){\small $b_1$} 
	\node[Nadjust=wh,Nmr=3](a7)(12,15){\small $b_0$} 
	\node[Nadjust=wh,Nmr=0](a8)(15,5){\small a} 
	\drawedge[ATnb=0,AHnb=1,ELside=r](b2,a3){\small 0} 
	\drawedge[ATnb=0,AHnb=1](a3,a4){\small 1} 
	\drawedge[ATnb=0,AHnb=1](a4,a5){\small 1} 	
	\drawedge[ATnb=0,AHnb=1](a5,a6){\small 1} 	
	\drawedge[ATnb=0,AHnb=1](a6,a7){\small 1} 	
	\drawedge[ATnb=0,AHnb=1](a7,a8){\small 1} 
	
	\node[Nadjust=wh,Nmr=3](b4)(47,45){\small $b_3$} 
	\node[Nadjust=wh,Nmr=3](b5)(50,35){\small $b_2$} 
	\node[Nadjust=wh,Nmr=3](b6)(53,25){\small $b_1$} 
	\node[Nadjust=wh,Nmr=3](b7)(56,15){\small $b_0$} 
	\node[Nadjust=wh,Nmr=0](b8)(53,5){\small d} 
	
	\node[Nadjust=wh,Nmr=3](c4)(37,45){\small $b_3$}

	\node[Nadjust=wh,Nmr=3](c5)(34,35){\small $b_2$} 
	\node[Nadjust=wh,Nmr=3](c6)(31,25){\small $b_1$} 
	\node[Nadjust=wh,Nmr=3,linewidth=0.5](c7)(28,15){\small $b_0$} 
	\node[Nadjust=wh,Nmr=0](c8)(25,5){\small b} 
	\node[Nadjust=wh,Nmr=0](c9)(31,5){\small c}

	\drawedge[ATnb=0,AHnb=1](b0,b1){\small 1} 
	\drawedge[ATnb=0,AHnb=1](b1,b2){\small 1}  
	\drawedge[ATnb=0,AHnb=1](b2,b3){\small 1}
	\drawedge[ATnb=0,AHnb=1](b3,b4){\small 1}
	\drawedge[ATnb=0,AHnb=1](b4,b5){\small 1} 
	\drawedge[ATnb=0,AHnb=1](b5,b6){\small 1}   
	\drawedge[ATnb=0,AHnb=1](b6,b7){\small 1} 
	\drawedge[ATnb=0,AHnb=1](b7,b8){\small 0} 
	\drawedge[ATnb=0,AHnb=1,ELside=r](b3,c4){\small 0}
	\drawedge[ATnb=0,AHnb=1,ELside=r](c4,c5){\small 0} 
	\drawedge[ATnb=0,AHnb=1,ELside=r](c5,c6){\small 0}   
	\drawedge[ATnb=0,AHnb=1,ELside=r](c6,c7){\small 0} 
	\drawedge[ATnb=0,AHnb=1,ELside=r](c7,c8){\small 0} 
	\drawedge[ATnb=0,AHnb=1,ELside=l](c7,c9){\small 1} 	
	
	\end{picture}	
	\caption{
	The search of $x$ in the compressed trie data structure.
	}
	\label{fig_pesquisaTrie}
	\end{center}
\end{figure*}
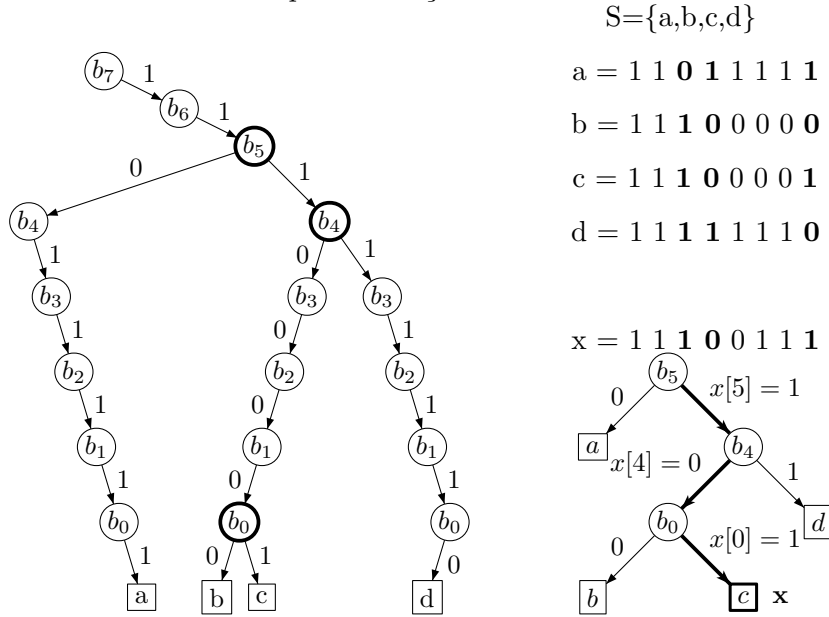

  \subsubsection{
   Computing $\rk{x}$
   }
  
  Suppose a set $S=(s_1,\ldots,s_t)$ inserted in a compressed trie. This section will describe how to compute $\rk{x}$ for a given key $x$. An initial search $s'=\trs{x}$ is computed. The $s'$ element has the same values than $x$ in the relevant bits. If $s'$ is equal to $x$ in the remaining bits, $\rk{x}=\rk{s'}+1$ and the rank is computed. 
In the other case, a new search will be performed. First, consider the bit $b'=\dlt{x}{s'}$.
  

\begin{lema} 
The bit $b'=\dlt{x}{s'}$ is the new relevant bit in the compressed trie with $S\cup\{x\}$ elements.
\end{lema}

The $\rk{x}$ calculus will be dived in two cases. In the first, the bit $b'$  of $x$ is 1, which means $x[b']=1$, while in the second case $x[b']=0$.


\begin{lema} 
The most significant bits of $x$ and $s'$ are equals. The first bit to diverge is $b’$. Consider the branch between $x$ and $s'$ in the trie with $S\cup\{x\}$. If  $x[b']=1$, the $x$ predecessor is the largest element in the $b'=0$ branch. If $x[b']=0$, the x successor is the smallest element in the branch $b'=1$.
\end{lema}

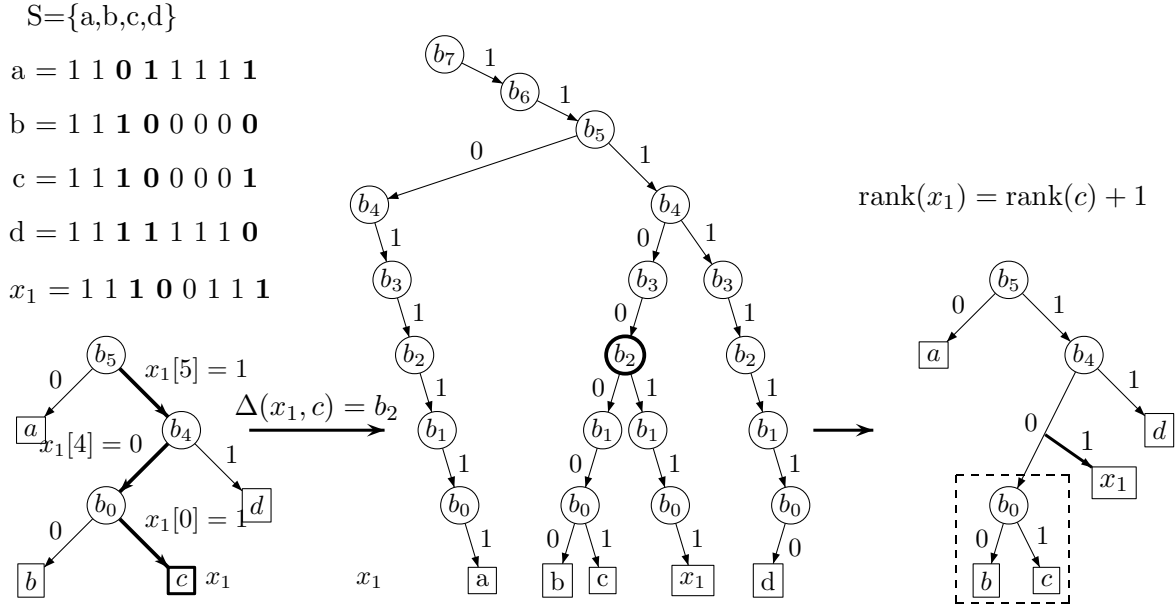
\begin{figure*}[htb]
	\begin{center}
	\setlength{\unitlength}{1mm}
\begin{picture}(160,80)(0,0)
	\put(128,2){\dashbox(15,17)}	
	\put(0,65){\begin{tabular}{r}
	  S=\{a,b,c,d\}\hspace{1.07cm}~\\
	  a = 1 1 {\bf 0 1} 1 1 1 {\bf 1}\\
	  b = 1 1 {\bf 1 0}  0 0 0 {\bf 0}\\
	  c = 1 1 {\bf 1 0} 0 0 0 {\bf 1}\\
	 d = 1 1 {\bf 1 1} 1 1 1 {\bf 0}\\
	\end{tabular}}
	\put(0,43){\begin{tabular}{r}
	 $x_1$ = 1 1 {\bf 1 0} 0 1 1 {\bf 1}			
	\end{tabular}}

	\node[Nframe=n,Nadjust=wh](x)(33,25){} 	
	\node[Nframe=n,Nadjust=wh](y)(53,25){} 	
	\drawedge[ATnb=0,AHnb=1,linewidth=.4,AHLength=2.5](x,y){$\dlt{x_1}{c}=b_2$} 	

	\node[Nadjust=wh,Nmr=3](fb2)(15,35){\small $b_5$} 
	\node[Nadjust=wh,Nmr=3](fb3)(25,25){\small $b_4$} 
	\node[Nadjust=wh,Nmr=3](fb7)(15,15){\small $b_0$} 
	\node[Nadjust=wh,Nmr=0](fc)(35,15){\small $d$} 
	\node[Nadjust=wh,Nmr=0](fb)(5,5){\small $b$} 
	\node[Nadjust=wh,Nmr=0,linewidth=.4](fd)(25,5){\small $c$} 
	\node[Nframe=n,Nadjust=wh,Nmr=0](x)(30,5){\small  $x_1$} 
	\node[Nadjust=wh,Nmr=0](fa)(5,25){\small $a$} 
	\drawedge[ATnb=0,AHnb=1,ELside=r,linewidth=.5,AHLength=2,ELpos=70](fb3,fb7){\small $x_1[4]=0$}
	\drawedge[ATnb=0,AHnb=1,ELside=l](fb3,fc){\small 1}
	\drawedge[ATnb=0,AHnb=1,ELside=l,linewidth=.5,AHLength=2,ELpos=70](fb7,fd){\small $x_1[0]=1$}
	\drawedge[ATnb=0,AHnb=1,ELside=r](fb7,fb){\small 0}
	\drawedge[ATnb=0,AHnb=1,ELside=l,linewidth=.5,AHLength=2,ELpos=70](fb2,fb3){\small $x_1[5]=1$}
	\drawedge[ATnb=0,AHnb=1,ELside=r](fb2,fa){\small 0}	
	
		\node[Nadjust=wh,Nmr=3](b0)(60,75){\small $b_7$} 
	\node[Nadjust=wh,Nmr=3](b1)(70,70){\small $b_6$} 
	\node[Nadjust=wh,Nmr=3](b2)(80,65){\small $b_5$} 
	\node[Nadjust=wh,Nmr=3](b3)(90,55){\small $b_4$} 
	\node[Nadjust=wh,Nmr=3](a3)(50,55){\small $b_4$} 
	\node[Nadjust=wh,Nmr=3](a4)(53,45){\small $b_3$} 
	\node[Nadjust=wh,Nmr=3](a5)(56,35){\small $b_2$} 
	\node[Nadjust=wh,Nmr=3](a6)(59,25){\small $b_1$} 
	\node[Nadjust=wh,Nmr=3](a7)(62,15){\small $b_0$} 
	\node[Nadjust=wh,Nmr=0](a8)(65,5){\small a} 
	\drawedge[ATnb=0,AHnb=1,ELside=r](b2,a3){\small 0} 
	\drawedge[ATnb=0,AHnb=1](a3,a4){\small 1} 
	\drawedge[ATnb=0,AHnb=1](a4,a5){\small 1} 	
	\drawedge[ATnb=0,AHnb=1](a5,a6){\small 1} 	
	\drawedge[ATnb=0,AHnb=1](a6,a7){\small 1} 	
	\drawedge[ATnb=0,AHnb=1](a7,a8){\small 1} 
	
	\node[Nadjust=wh,Nmr=3](b4)(97,45){\small $b_3$} 
	\node[Nadjust=wh,Nmr=3](b5)(100,35){\small $b_2$} 
	\node[Nadjust=wh,Nmr=3](b6)(103,25){\small $b_1$} 
	\node[Nadjust=wh,Nmr=3](b7)(106,15){\small $b_0$} 
	\node[Nadjust=wh,Nmr=0](b8)(103,5){\small d} 
	
	\node[Nadjust=wh,Nmr=3](c4)(87,45){\small $b_3$}

	\node[Nadjust=wh,Nmr=3,linewidth=.5](c5)(84,35){\small $b_2$} 
	\node[Nadjust=wh,Nmr=3](c6)(81,25){\small $b_1$}
	\node[Nadjust=wh,Nmr=3](c7)(78,15){\small $b_0$} 
	\node[Nadjust=wh,Nmr=0](c8)(75,5){\small b} 
	\node[Nadjust=wh,Nmr=0](c9)(81,5){\small c}

	\node[Nadjust=wh,Nmr=3](x6)(87,25){\small $b_1$} 
	\node[Nadjust=wh,Nmr=3](x7)(90,15){\small $b_0$} 
	\node[Nadjust=wh,Nmr=0](x8)(93,5){\small $x_1$} 	
         \drawedge[ATnb=0,AHnb=1](c5,x6){\small 1} 
          \drawedge[ATnb=0,AHnb=1](x6,x7){\small 1} 
           \drawedge[ATnb=0,AHnb=1](x7,x8){\small 1} 
	
	\drawedge[ATnb=0,AHnb=1](b0,b1){\small 1} 
	\drawedge[ATnb=0,AHnb=1](b1,b2){\small 1}  
	\drawedge[ATnb=0,AHnb=1](b2,b3){\small 1}
	\drawedge[ATnb=0,AHnb=1](b3,b4){\small 1}
	\drawedge[ATnb=0,AHnb=1](b4,b5){\small 1} 
	\drawedge[ATnb=0,AHnb=1](b5,b6){\small 1}   
	\drawedge[ATnb=0,AHnb=1](b6,b7){\small 1} 
	\drawedge[ATnb=0,AHnb=1](b7,b8){\small 0} 
	\drawedge[ATnb=0,AHnb=1,ELside=r](b3,c4){\small 0}
	\drawedge[ATnb=0,AHnb=1,ELside=r](c4,c5){\small 0} 
	\drawedge[ATnb=0,AHnb=1,ELside=r](c5,c6){\small 0}   
	\drawedge[ATnb=0,AHnb=1,ELside=r](c6,c7){\small 0} 
	\drawedge[ATnb=0,AHnb=1,ELside=r](c7,c8){\small 0} 
	\drawedge[ATnb=0,AHnb=1,ELside=l](c7,c9){\small 1}

	\node[Nframe=n,Nadjust=wh](x)(108,25){} 	
	\node[Nframe=n,Nadjust=wh](y)(118,25){} 	
	\drawedge[ATnb=0,AHnb=1,linewidth=.4,AHLength=2.5](x,y){}

	\node[Nframe=n,Nadjust=wh](z)(139,25){} 	
	\node[Nmr=0,Nadjust=wh](w)(149,18){$x_1$} 	
	\drawedge[ATnb=0,AHnb=1,linewidth=.4](z,w){1}

	\node[Nadjust=wh,Nmr=3](gb2)(135,45){\small $b_5$} 
	\node[Nadjust=wh,Nmr=3](gb3)(145,35){\small $b_4$} 
	\node[Nadjust=wh,Nmr=3](gb7)(135,15){\small $b_0$} 
	\node[Nadjust=wh,Nmr=0](gc)(155,25){\small $d$} 
	\node[Nadjust=wh,Nmr=0](gb)(132,5){\small $b$} 
	\node[Nadjust=wh,Nmr=0](gd)(140,5){\small $c$} 
	\node[Nframe=n,Nadjust=wh,Nmr=0](gx)(50,5){\small $x_1$} 
	\node[Nadjust=wh,Nmr=0](ga)(125,35){\small $a$} 
	\drawedge[ATnb=0,AHnb=1,ELside=r](gb3,gb7){\small 0}
	\drawedge[ATnb=0,AHnb=1,ELside=l](gb3,gc){\small 1}
	\drawedge[ATnb=0,AHnb=1,ELside=l](gb7,gd){\small 1}
	\drawedge[ATnb=0,AHnb=1,ELside=r](gb7,gb){\small 0}
	\drawedge[ATnb=0,AHnb=1,ELside=l](gb2,gb3){\small 1}
	\drawedge[ATnb=0,AHnb=1,ELside=r](gb2,ga){\small 0}	
	\put(115,55){$\rk{x_1}=\rk{c}+1$}

	\end{picture}		
	\setlength{\unitlength}{1mm}
	\caption{
	Computing $\rk{x_1}$.
	}
	\label{rk}
	\end{center}
\end{figure*}

\textbf{Case a ($x[b']=1$)} Figure~\ref{rk} has an example in which the predecessor of $x$ is the largest element in the subtree highlighted.


A second search is needed to compute $\rk{x}$. From the most significant bit to the relevant bit $b'$, the compressed trie search uses the bits of $x_4$. Starting from $b'$, the search looks for the largest element in the subtree, i.e., the search will down the tree always to the right branch in direction of the largest element.

A new search key $x'$ will be computed to obtain such behavior in the following way:


\begin{tabular}{rr}
& $x =x_{w-1}x_{w-2} \ldots x_2x_1x_0$\\
OR & $ 1~ 1 ~1 ~1 $  \\ \hline
&$x' = x_{w-1}x_{w-2} \ldots 1~1~1~1$
\end{tabular}

\begin{figure*}[htb]
	\begin{center}
	\setlength{\unitlength}{1mm}
\begin{picture}(160,80)(0,0)
	\put(0,65){\begin{tabular}{cc}
	  S=\{a,b,c,d\}~~~~\\
	  a = 1 1 {\bf 0 1} 1 1 1 {\bf 1}\\
	  b = 1 1 {\bf 1 0}  0 0 0 {\bf 0}\\
	  c = 1 1 {\bf 1 0} 0 0 0 {\bf 1}\\
	  d = 1 1 {\bf 1 1} 1 1 1 {\bf 0}\\
	\end{tabular}}
	\put(0,43){\begin{tabular}{cc}
	 $x_2$ = 1 1 {\bf 1 0} 1 0 0 {\bf 0}			
	\end{tabular}}

	\node[Nframe=n,Nadjust=wh](x)(33,25){} 	
	\node[Nframe=n,Nadjust=wh](y)(53,25){} 	
	\drawedge[ATnb=0,AHnb=1,linewidth=.4,AHLength=2.5](x,y){$\dlt{x_2}{b}=b_3$} 	

	\node[Nadjust=wh,Nmr=3](fb2)(15,35){\small $b_5$} 
	\node[Nadjust=wh,Nmr=3](fb3)(25,25){\small $b_4$} 
	\node[Nadjust=wh,Nmr=3](fb7)(15,15){\small $b_0$} 
	\node[Nadjust=wh,Nmr=0](fc)(35,15){\small $d$} 
	\node[Nadjust=wh,Nmr=0,linewidth=.4](fb)(5,5){\small $b$} 
	\node[Nadjust=wh,Nmr=0](fd)(25,5){\small $c$} 
	\node[Nframe=n,Nadjust=wh,Nmr=0](x)(0,5){\small  $x_2$} 
	\node[Nadjust=wh,Nmr=0](fa)(5,25){\small $a$} 
	\drawedge[ATnb=0,AHnb=1,ELside=r,linewidth=.4,AHLength=2.5,ELpos=70](fb3,fb7){\small $x_2[4]=0$}
	\drawedge[ATnb=0,AHnb=1,ELside=l](fb3,fc){\small 1}
	\drawedge[ATnb=0,AHnb=1,ELside=l](fb7,fd){\small $1$}
	\drawedge[ATnb=0,AHnb=1,ELside=r,linewidth=.4,AHLength=2.5,ELpos=70](fb7,fb){\small $x_2[0]=0$}
	\drawedge[ATnb=0,AHnb=1,ELside=l,linewidth=.4,AHLength=2.5,ELpos=70](fb2,fb3){\small $x_2[5]=1$}
	\drawedge[ATnb=0,AHnb=1,ELside=r](fb2,fa){\small 0}	
	
		\node[Nadjust=wh,Nmr=3](b0)(60,75){\small $b_7$} 
	\node[Nadjust=wh,Nmr=3](b1)(70,70){\small $b_6$} 
	\node[Nadjust=wh,Nmr=3](b2)(80,65){\small $b_5$} 
	\node[Nadjust=wh,Nmr=3](b3)(90,55){\small $b_4$} 
	\node[Nadjust=wh,Nmr=3](a3)(50,55){\small $b_4$} 
	\node[Nadjust=wh,Nmr=3](a4)(53,45){\small $b_3$} 
	\node[Nadjust=wh,Nmr=3](a5)(56,35){\small $b_2$} 
	\node[Nadjust=wh,Nmr=3](a6)(59,25){\small $b_1$} 
	\node[Nadjust=wh,Nmr=3](a7)(62,15){\small $b_0$} 
	\node[Nadjust=wh,Nmr=0](a8)(65,5){\small a} 
	\drawedge[ATnb=0,AHnb=1,ELside=r](b2,a3){\small 0} 
	\drawedge[ATnb=0,AHnb=1](a3,a4){\small 1} 
	\drawedge[ATnb=0,AHnb=1](a4,a5){\small 1} 	
	\drawedge[ATnb=0,AHnb=1](a5,a6){\small 1} 	
	\drawedge[ATnb=0,AHnb=1](a6,a7){\small 1} 	
	\drawedge[ATnb=0,AHnb=1](a7,a8){\small 1} 
	
	\node[Nadjust=wh,Nmr=3](b4)(97,45){\small $b_3$} 
	\node[Nadjust=wh,Nmr=3](b5)(100,35){\small $b_2$} 
	\node[Nadjust=wh,Nmr=3](b6)(103,25){\small $b_1$} 
	\node[Nadjust=wh,Nmr=3](b7)(106,15){\small $b_0$} 
	\node[Nadjust=wh,Nmr=0](b8)(103,5){\small d} 
	
	\node[Nadjust=wh,Nmr=3,linewidth=.5](c4)(87,45){\small $b_3$}

	\node[Nadjust=wh,Nmr=3](c5)(84,35){\small $b_2$} 
	\node[Nadjust=wh,Nmr=3](c6)(81,25){\small $b_1$}
	\node[Nadjust=wh,Nmr=3](c7)(78,15){\small $b_0$} 
	\node[Nadjust=wh,Nmr=0](c8)(75,5){\small b} 
	\node[Nadjust=wh,Nmr=0](c9)(81,5){\small c}

	\node[Nadjust=wh,Nmr=3](x5)(90,35){\small $b_2$} 
	\node[Nadjust=wh,Nmr=3](x6)(92,25){\small $b_1$} 
	\node[Nadjust=wh,Nmr=3](x7)(94,15){\small $b_0$} 
	\node[Nadjust=wh,Nmr=0](x8)(96,5){\small  $x_2$} 	
         \drawedge[ATnb=0,AHnb=1](x5,x6){\small 1} 
          \drawedge[ATnb=0,AHnb=1](x6,x7){\small 1} 
           \drawedge[ATnb=0,AHnb=1](x7,x8){\small 1} 
	
	\drawedge[ATnb=0,AHnb=1](c4,x5){\small 1} 
	
	\drawedge[ATnb=0,AHnb=1](b0,b1){\small 1} 
	\drawedge[ATnb=0,AHnb=1](b1,b2){\small 1}  
	\drawedge[ATnb=0,AHnb=1](b2,b3){\small 1}
	\drawedge[ATnb=0,AHnb=1](b3,b4){\small 1}
	\drawedge[ATnb=0,AHnb=1](b4,b5){\small 1} 
	\drawedge[ATnb=0,AHnb=1](b5,b6){\small 1}   
	\drawedge[ATnb=0,AHnb=1](b6,b7){\small 1} 
	\drawedge[ATnb=0,AHnb=1](b7,b8){\small 0} 
	\drawedge[ATnb=0,AHnb=1,ELside=r](b3,c4){\small 0}
	\drawedge[ATnb=0,AHnb=1,ELside=r](c4,c5){\small 0} 
	\drawedge[ATnb=0,AHnb=1,ELside=r](c5,c6){\small 0}   
	\drawedge[ATnb=0,AHnb=1,ELside=r](c6,c7){\small 0} 
	\drawedge[ATnb=0,AHnb=1,ELside=r](c7,c8){\small 0} 
	\drawedge[ATnb=0,AHnb=1,ELside=l](c7,c9){\small 1}

	\node[Nframe=n,Nadjust=wh](x)(108,25){} 	
	\node[Nframe=n,Nadjust=wh](y)(118,25){} 	
	\drawedge[ATnb=0,AHnb=1,linewidth=.4,AHLength=2.5](x,y){}


	\node[Nadjust=wh,Nmr=3](gb2)(135,45){\small $b_5$} 
	\node[Nadjust=wh,Nmr=3](gb3)(145,35){\small $b_4$} 
	\node[Nadjust=wh,Nmr=3](gb7)(135,23){\small $b_0$} 
	\node[Nadjust=wh,Nmr=0](gc)(155,25){\small $d$} 
	\node[Nadjust=wh,Nmr=0](gb)(132,10){\small $b$} 
	\node[Nadjust=wh,Nmr=0,linewidth=.5](gd)(140,10){\small $c$} 
	\node[Nframe=n,Nadjust=wh,Nmr=0](gx)(50,10){\small  x} 
	\node[Nadjust=wh,Nmr=0](ga)(125,35){\small $a$} 
	\drawedge[ATnb=0,AHnb=1,ELside=r,linewidth=.5,ELpos=70](gb3,gb7){\small $x_2'[4]=0$}
	\drawedge[ATnb=0,AHnb=1,ELside=l](gb3,gc){\small 1}
	\drawedge[ATnb=0,AHnb=1,ELside=l,linewidth=.5,ELpos=70](gb7,gd){\small $x_2'[0]=1$}
	\drawedge[ATnb=0,AHnb=1,ELside=r](gb7,gb){\small 0}
	\drawedge[ATnb=0,AHnb=1,ELside=l,linewidth=.5,ELpos=70](gb2,gb3){\small $x_2'[5]=1$}
	\drawedge[ATnb=0,AHnb=1,ELside=r](gb2,ga){\small 0}	
	\put(115,60){\begin{tabular}{cc}
	$x_2'$&$= x_2 ~OR~ 1111$ \\
	&$=1 1  {\bf 1 0}  ~ 1 1 1 {\bf 1}$
	\end{tabular}
	}
	\put(120,0){$\rk{x_2}=\rk{c}+1$}

	\end{picture}		
	\setlength{\unitlength}{1mm}
	\caption{
	Computing $\rk{x_2}$ after the second search in the compressed trie. 
	}
	\label{rkd}
	\end{center}
\end{figure*}
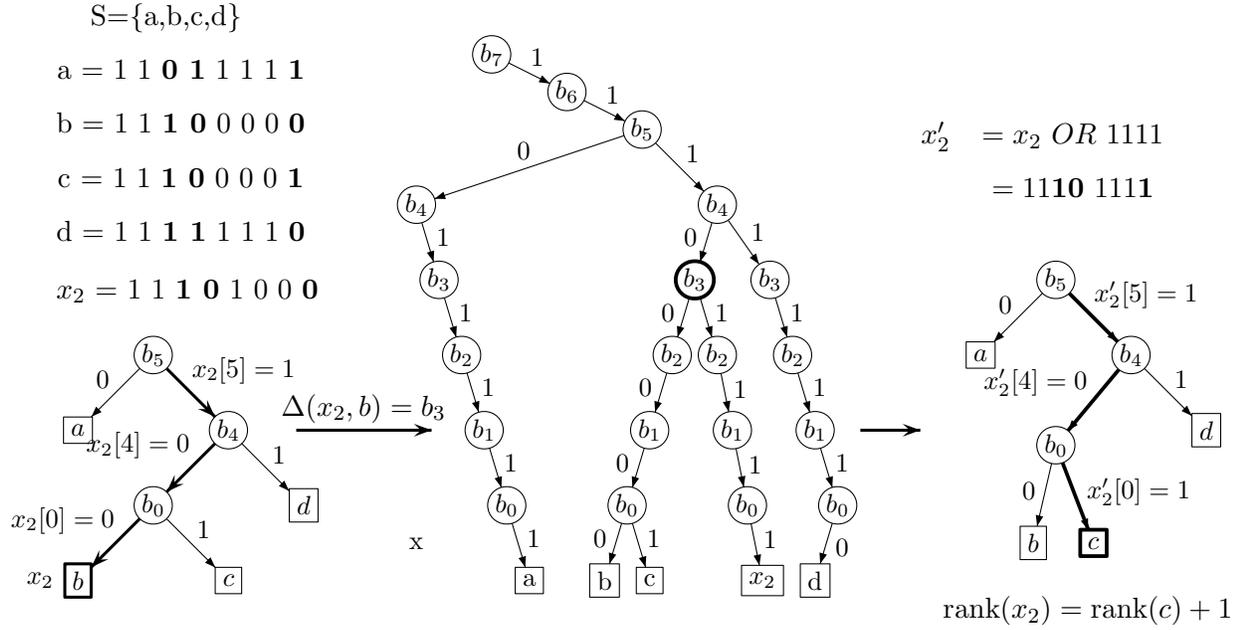

The number of 1’s at the end of $x'$ is $b'$. Such mask can be computed in $O(1)$ as $2^{b'+1}-1$.
When a bit of $x$ is replaced by 1 from $b'$ to $b_0$, the new search key will find $x$ predecessor. Let $s''=\trs{x'}$. Then $\rk{x}=\rk{s''}+1$. Figure~\ref{rkd} has a sample.


\textbf{Case b ($x[b']=0$) } In the Figure~\ref{rkb} example, the $x_3$ successor is the smaller element in the highlighted subtree. A second search is performed to compute $\rk{x_3}$. From the most significant bit to the relevant bit $b'$, the compressed trie search uses the bits of $x$. Starting from $b'$, the search looks for the smallest element in the subtree, i.e., the search will down the tree always to the left branch in direction of the smallest element.
%

%
\begin{figure*}[htb]
	\begin{center}
	\setlength{\unitlength}{.9mm}
\begin{picture}(180,79)(0,0)
	\put(141,2){\dashbox(37,48)}	
	\put(0,65){\begin{tabular}{r}
	  S=\{a,b,c,d\}\hspace{1.07cm}~\\
	  a = 1 1 {\bf 0 1} 1 1 1 {\bf 1}\\
	  b = 1 1 {\bf 1 0}  0 0 0 {\bf 0}\\
	  c = 1 1 {\bf 1 0} 0 0 0 {\bf 1}\\
	  d = 1 1 {\bf 1 1} 1 1 1 {\bf 0}\\			
	\end{tabular}}
	\put(0,43){\begin{tabular}{r}	  		
	 $x_3$ = 0 1 {\bf 1 0} 0 1 1 {\bf 1}			
	\end{tabular}}

	\node[Nframe=n,Nadjust=wh](x)(33,25){} 	
	\node[Nframe=n,Nadjust=wh](y)(53,25){} 	
	\drawedge[ATnb=0,AHnb=1,linewidth=.4,AHLength=2.5](x,y){$\dlt{x_3}{c}=b_7$} 	

	\node[Nadjust=wh,Nmr=3](fb2)(15,35){\small $b_5$} 
	\node[Nadjust=wh,Nmr=3](fb3)(25,25){\small $b_4$} 
	\node[Nadjust=wh,Nmr=3](fb7)(15,15){\small $b_0$} 
	\node[Nadjust=wh,Nmr=0](fc)(35,15){\small $d$} 
	\node[Nadjust=wh,Nmr=0](fb)(5,5){\small $b$} 
	\node[Nadjust=wh,Nmr=0,linewidth=.4](fd)(25,5){\small $c$} 
	\node[Nframe=n,Nadjust=wh,Nmr=0](x)(30,5){\small  $x_3$} 
	\node[Nadjust=wh,Nmr=0](fa)(5,25){\small $a$} 
	\drawedge[ATnb=0,AHnb=1,ELside=r,linewidth=.5,AHLength=2,ELpos=90](fb3,fb7){\small $x_3[4]=0$}
	\drawedge[ATnb=0,AHnb=1,ELside=l](fb3,fc){\small 1}
	\drawedge[ATnb=0,AHnb=1,ELside=l,linewidth=.5,AHLength=2,ELpos=100](fb7,fd){\small $x_3[0]=1$}
	\drawedge[ATnb=0,AHnb=1,ELside=r](fb7,fb){\small 0}
	\drawedge[ATnb=0,AHnb=1,ELside=l,linewidth=.5,AHLength=2,ELpos=70](fb2,fb3){\small $x_3[5]=1$}
	\drawedge[ATnb=0,AHnb=1,ELside=r](fb2,fa){\small 0}

	\node[Nadjust=wh,Nmr=3](bb6)(52,70){\small $b_6$}
	\node[Nadjust=wh,Nmr=3](bb5)(60,65){\small $b_5$} 
	\node[Nadjust=wh,Nmr=3](bb4)(63,55){\small $b_4$} 
	\node[Nadjust=wh,Nmr=3](bb3)(60,45){\small $b_3$} 
	\node[Nadjust=wh,Nmr=3](bb2)(57,35){\small $b_2$} 
	\node[Nadjust=wh,Nmr=3](bb1)(60,25){\small $b_1$} 
	\node[Nadjust=wh,Nmr=3](bb0)(63,15){\small $b_0$} 
	\node[Nadjust=wh,Nmr=0](x3)(66,5){\small  $x_3$}	
	\drawedge[ATnb=0,AHnb=1](bb6,bb5){\small 1} 
	\drawedge[ATnb=0,AHnb=1](bb5,bb4){\small 1} 
	\drawedge[ATnb=0,AHnb=1,ELside=r](bb4,bb3){\small 0} 
	\drawedge[ATnb=0,AHnb=1,ELside=r](bb3,bb2){\small 0} 	
	\drawedge[ATnb=0,AHnb=1](bb2,bb1){\small 1} 	
	\drawedge[ATnb=0,AHnb=1](bb1,bb0){\small 1} 	
	\drawedge[ATnb=0,AHnb=1](bb0,x3){\small 1} 		
		\node[Nadjust=wh,Nmr=3,linewidth=.4](b0)(80,75){\small $b_7$} 
		\drawedge[ATnb=0,AHnb=1,ELside=r](b0,bb6){\small 0} 
		
	\node[Nadjust=wh,Nmr=3](b1)(90,70){\small $b_6$} 
	\node[Nadjust=wh,Nmr=3](b2)(100,65){\small $b_5$} 
	\node[Nadjust=wh,Nmr=3](b3)(110,55){\small $b_4$} 
	\node[Nadjust=wh,Nmr=3](a3)(72,55){\small $b_4$} 
	\node[Nadjust=wh,Nmr=3](a4)(75,45){\small $b_3$} 
	\node[Nadjust=wh,Nmr=3](a5)(77,35){\small $b_2$} 
	\node[Nadjust=wh,Nmr=3](a6)(79,25){\small $b_1$} 
	\node[Nadjust=wh,Nmr=3](a7)(82,15){\small $b_0$} 
	\node[Nadjust=wh,Nmr=0](a8)(85,5){\small a} 
	\drawedge[ATnb=0,AHnb=1,ELside=r](b2,a3){\small 0} 
	\drawedge[ATnb=0,AHnb=1](a3,a4){\small 1} 
	\drawedge[ATnb=0,AHnb=1](a4,a5){\small 1} 	
	\drawedge[ATnb=0,AHnb=1](a5,a6){\small 1} 	
	\drawedge[ATnb=0,AHnb=1](a6,a7){\small 1} 	
	\drawedge[ATnb=0,AHnb=1](a7,a8){\small 1} 
	
	\node[Nadjust=wh,Nmr=3](b4)(117,45){\small $b_3$} 
	\node[Nadjust=wh,Nmr=3](b5)(120,35){\small $b_2$} 
	\node[Nadjust=wh,Nmr=3](b6)(123,25){\small $b_1$} 
	\node[Nadjust=wh,Nmr=3](b7)(126,15){\small $b_0$} 
	\node[Nadjust=wh,Nmr=0](b8)(123,5){\small d} 
	
	\node[Nadjust=wh,Nmr=3](c4)(107,45){\small $b_3$}

	\node[Nadjust=wh,Nmr=3](c5)(104,35){\small $b_2$} 
	\node[Nadjust=wh,Nmr=3](c6)(101,25){\small $b_1$}
	\node[Nadjust=wh,Nmr=3](c7)(98,15){\small $b_0$} 
	\node[Nadjust=wh,Nmr=0](c8)(95,5){\small b} 
	\node[Nadjust=wh,Nmr=0](c9)(101,5){\small c}

	\node[Nadjust=wh,Nmr=3](x6)(107,25){\small $b_1$} 
	\node[Nadjust=wh,Nmr=3](x7)(110,15){\small $b_0$} 
	\node[Nadjust=wh,Nmr=0](x8)(113,5){\small  $x_3$} 	
         \drawedge[ATnb=0,AHnb=1](c5,x6){\small 1} 
          \drawedge[ATnb=0,AHnb=1](x6,x7){\small 1} 
           \drawedge[ATnb=0,AHnb=1](x7,x8){\small 1} 
	
	\drawedge[ATnb=0,AHnb=1](b0,b1){\small 1} 
	\drawedge[ATnb=0,AHnb=1](b1,b2){\small 1}  
	\drawedge[ATnb=0,AHnb=1](b2,b3){\small 1}
	\drawedge[ATnb=0,AHnb=1](b3,b4){\small 1}
	\drawedge[ATnb=0,AHnb=1](b4,b5){\small 1} 
	\drawedge[ATnb=0,AHnb=1](b5,b6){\small 1}   
	\drawedge[ATnb=0,AHnb=1](b6,b7){\small 1} 
	\drawedge[ATnb=0,AHnb=1](b7,b8){\small 0} 
	\drawedge[ATnb=0,AHnb=1,ELside=r](b3,c4){\small 0}
	\drawedge[ATnb=0,AHnb=1,ELside=r](c4,c5){\small 0} 
	\drawedge[ATnb=0,AHnb=1,ELside=r](c5,c6){\small 0}   
	\drawedge[ATnb=0,AHnb=1,ELside=r](c6,c7){\small 0} 
	\drawedge[ATnb=0,AHnb=1,ELside=r](c7,c8){\small 0} 
	\drawedge[ATnb=0,AHnb=1,ELside=l](c7,c9){\small 1}

	\node[Nframe=n,Nadjust=wh](x)(128,25){} 	
	\node[Nframe=n,Nadjust=wh](y)(138,25){} 	
	\drawedge[ATnb=0,AHnb=1,linewidth=.4,AHLength=2.5](x,y){}

	\node[Nadjust=wh,Nmr=3](gb2)(155,45){\small $b_5$} 
	\node[Nadjust=wh,Nmr=3](gb3)(165,35){\small $b_4$} 
	\node[Nadjust=wh,Nmr=3](gb7)(155,15){\small $b_0$} 
	\node[Nadjust=wh,Nmr=0](gc)(175,25){\small $d$} 
	\node[Nadjust=wh,Nmr=0](gb)(152,5){\small $b$} 
	\node[Nadjust=wh,Nmr=0](gd)(160,5){\small $c$} 
	\node[Nadjust=wh,Nmr=0](ga)(145,35){\small $a$} 
	\drawedge[ATnb=0,AHnb=1,ELside=r](gb3,gb7){\small 0}
	\drawedge[ATnb=0,AHnb=1,ELside=l](gb3,gc){\small 1}
	\drawedge[ATnb=0,AHnb=1,ELside=l](gb7,gd){\small 1}
	\drawedge[ATnb=0,AHnb=1,ELside=r](gb7,gb){\small 0}
	\drawedge[ATnb=0,AHnb=1,ELside=l](gb2,gb3){\small 1}
	\drawedge[ATnb=0,AHnb=1,ELside=r](gb2,ga){\small 0}	

	\node[Nframe=n,Nadjust=wh](z2)(145,65){} 	
	\node[Nframe=n,Nadjust=wh](z)(147,63){} 	
	\node[Nmr=0,Nadjust=wh](w)(135,55){ $x_3$} 	
	\drawedge[ATnb=0,AHnb=1,linewidth=.4,ELside=r](z,w){0} 	
	\drawedge[ATnb=0,AHnb=1,ELside=l](z2,gb2){\small 1}

	\end{picture}		
	\setlength{\unitlength}{1mm}
	\caption{
	Computing $\rk{x_3}$ after the second search in the compressed trie.}
	\label{rkb}
	\end{center}
\end{figure*}

A new search key $x’$ will be computed to obtain such behavior in the following way:


\begin{tabular}{rr}
& $x =x_{w-1}x_{w-2} \ldots x_2x_1x_0$\\
AND & $ 1~ 1 ~1 ~1\ldots0~ 0 ~0 ~0 $  \\ \hline
&$x' = x_{w-1}x_{w-2} \ldots 0~0~0~0$
\end{tabular}

The number of zeros at the end of $x’$ is $b’$. When a bit of $x$ is replaced by zero from $b'$ to $b_0$, the new search key will find $x$ successor. Let $s''=\trs{x'}$. Then $\rk{x}=\rk{s''}$. Figure~\ref{rkt} has a sample.


\begin{figure*}[htb]
	\begin{center}
	\setlength{\unitlength}{1mm}
\begin{picture}(160,80)(0,0)
	\put(0,60){\begin{tabular}{rl}
	 & S=\{a,b,c,d\}\\
	  a\hspace{-.3cm}&\hspace{-.3cm}=1 1 {\bf 0 1} 1 1 1 {\bf 1}\\
	  b\hspace{-.3cm}&\hspace{-.3cm}=1 1 {\bf 1 0}  0 1 0 {\bf 0}\\
	  c\hspace{-.3cm}&\hspace{-.3cm}=1 1 {\bf 1 0} 0 1 0 {\bf 1}\\
	  d\hspace{-.3cm}&\hspace{-.3cm}=1 1 {\bf 1 1} 1 1 1 {\bf 0}\\
			
	 $x_4$\hspace{-.3cm}&\hspace{-.3cm}=1 1 {\bf 1 0} 0 0 0 {\bf 1}			
	\end{tabular}}

	\node[Nframe=n,Nadjust=wh](x)(33,25){} 	
	\node[Nframe=n,Nadjust=wh](y)(53,25){} 	
	\drawedge[ATnb=0,AHnb=1,linewidth=.4,AHLength=2.5](x,y){$\dlt{x_4}{c}=b_2$} 	

	\node[Nadjust=wh,Nmr=3](fb2)(15,35){\small $b_5$} 
	\node[Nadjust=wh,Nmr=3](fb3)(25,25){\small $b_4$} 
	\node[Nadjust=wh,Nmr=3](fb7)(15,15){\small $b_0$} 
	\node[Nadjust=wh,Nmr=0](fc)(35,15){\small $d$} 
	\node[Nadjust=wh,Nmr=0](fb)(5,5){\small $b$} 
	\node[Nadjust=wh,Nmr=0,linewidth=.4](fd)(25,5){\small $c$} 
	\node[Nframe=n,Nadjust=wh,Nmr=0](x)(0,5){\small  $x_4$} 
	\node[Nadjust=wh,Nmr=0](fa)(5,25){\small $a$} 
	\drawedge[ATnb=0,AHnb=1,ELside=r,linewidth=.4,AHLength=2.5,ELpos=90](fb3,fb7){\small $x_4[4]=0$}
	\drawedge[ATnb=0,AHnb=1,ELside=l](fb3,fc){\small 1}
	\drawedge[ATnb=0,AHnb=1,ELside=l,linewidth=.4,AHLength=2.5,ELpos=90](fb7,fd){\small $x_4[0]=1$}
	\drawedge[ATnb=0,AHnb=1,ELside=r](fb7,fb){\small $0$}
	\drawedge[ATnb=0,AHnb=1,ELside=l,linewidth=.4,AHLength=2.5,ELpos=70](fb2,fb3){\small $x_4[5]=1$}
	\drawedge[ATnb=0,AHnb=1,ELside=r](fb2,fa){\small 0}	
	
		\node[Nadjust=wh,Nmr=3](b0)(60,75){\small $b_7$} 
	\node[Nadjust=wh,Nmr=3](b1)(70,70){\small $b_6$} 
	\node[Nadjust=wh,Nmr=3](b2)(80,65){\small $b_5$} 
	\node[Nadjust=wh,Nmr=3](b3)(90,55){\small $b_4$} 
	\node[Nadjust=wh,Nmr=3](a3)(50,55){\small $b_4$} 
	\node[Nadjust=wh,Nmr=3](a4)(53,45){\small $b_3$} 
	\node[Nadjust=wh,Nmr=3](a5)(56,35){\small $b_2$} 
	\node[Nadjust=wh,Nmr=3](a6)(59,25){\small $b_1$} 
	\node[Nadjust=wh,Nmr=3](a7)(62,15){\small $b_0$} 
	\node[Nadjust=wh,Nmr=0](a8)(65,5){\small a} 
	\drawedge[ATnb=0,AHnb=1,ELside=r](b2,a3){\small 0} 
	\drawedge[ATnb=0,AHnb=1](a3,a4){\small 1} 
	\drawedge[ATnb=0,AHnb=1](a4,a5){\small 1} 	
	\drawedge[ATnb=0,AHnb=1](a5,a6){\small 1} 	
	\drawedge[ATnb=0,AHnb=1](a6,a7){\small 1} 	
	\drawedge[ATnb=0,AHnb=1](a7,a8){\small 1} 
	
	\node[Nadjust=wh,Nmr=3](b4)(97,45){\small $b_3$} 
	\node[Nadjust=wh,Nmr=3](b5)(100,35){\small $b_2$} 
	\node[Nadjust=wh,Nmr=3](b6)(103,25){\small $b_1$} 
	\node[Nadjust=wh,Nmr=3](b7)(106,15){\small $b_0$} 
	\node[Nadjust=wh,Nmr=0](b8)(103,5){\small d} 
	
	\node[Nadjust=wh,Nmr=3](c4)(87,45){\small $b_3$}

	\node[Nadjust=wh,Nmr=3,linewidth=.5](c5)(84,35){\small $b_2$} 
	\node[Nadjust=wh,Nmr=3](c6)(93,25){\small $b_1$}
	\node[Nadjust=wh,Nmr=3](c7)(87,15){\small $b_0$} 
	\node[Nadjust=wh,Nmr=0](c8)(84,5){\small b} 
	\node[Nadjust=wh,Nmr=0](c9)(90,5){\small c}

	\node[Nadjust=wh,Nmr=3](x6)(79,25){\small $b_1$} 
	\node[Nadjust=wh,Nmr=3](x7)(75,15){\small $b_0$} 
	\node[Nadjust=wh,Nmr=0](x8)(75,5){\small  $x_4$} 	
         \drawedge[ATnb=0,AHnb=1,ELside=r](c5,x6){\small 0} 
          \drawedge[ATnb=0,AHnb=1,ELside=r](x6,x7){\small 0} 
           \drawedge[ATnb=0,AHnb=1](x7,x8){\small 1} 
%
	
	\drawedge[ATnb=0,AHnb=1](b0,b1){\small 1} 
	\drawedge[ATnb=0,AHnb=1](b1,b2){\small 1}  
	\drawedge[ATnb=0,AHnb=1](b2,b3){\small 1}
	\drawedge[ATnb=0,AHnb=1](b3,b4){\small 1}
	\drawedge[ATnb=0,AHnb=1](b4,b5){\small 1} 
	\drawedge[ATnb=0,AHnb=1](b5,b6){\small 1}   
	\drawedge[ATnb=0,AHnb=1](b6,b7){\small 1} 
	\drawedge[ATnb=0,AHnb=1](b7,b8){\small 0} 
	\drawedge[ATnb=0,AHnb=1,ELside=r](b3,c4){\small 0}
	\drawedge[ATnb=0,AHnb=1,ELside=r](c4,c5){\small 0} 
	\drawedge[ATnb=0,AHnb=1,ELside=l](c5,c6){\small 1}   
	\drawedge[ATnb=0,AHnb=1,ELside=r](c6,c7){\small 0} 
	\drawedge[ATnb=0,AHnb=1,ELside=r](c7,c8){\small 0} 
	\drawedge[ATnb=0,AHnb=1,ELside=l](c7,c9){\small 1}

	\node[Nframe=n,Nadjust=wh](x)(108,25){} 	
	\node[Nframe=n,Nadjust=wh](y)(118,25){} 	
	\drawedge[ATnb=0,AHnb=1,linewidth=.4,AHLength=2.5](x,y){}


	\node[Nadjust=wh,Nmr=3](gb2)(135,45){\small $b_5$} 
	\node[Nadjust=wh,Nmr=3](gb3)(145,35){\small $b_4$} 
	\node[Nadjust=wh,Nmr=3](gb7)(135,23){\small $b_0$} 
	\node[Nadjust=wh,Nmr=0](gc)(155,25){\small $d$} 
	\node[Nadjust=wh,Nmr=0,,linewidth=.5](gb)(132,10){\small $b$} 
	\node[Nadjust=wh,Nmr=0](gd)(140,10){\small $c$} 
	\node[Nframe=n,Nadjust=wh,Nmr=0](gx)(50,10){\small  x} 
	\node[Nadjust=wh,Nmr=0](ga)(125,35){\small $a$} 
	\drawedge[ATnb=0,AHnb=1,ELside=r,linewidth=.5,ELpos=70](gb3,gb7){\small $x_4'[4]=0$}
	\drawedge[ATnb=0,AHnb=1,ELside=l](gb3,gc){\small 1}
	\drawedge[ATnb=0,AHnb=1,ELside=l](gb7,gd){\small $1$}
	\drawedge[ATnb=0,AHnb=1,ELside=r,linewidth=.5,ELpos=70](gb7,gb){\small $x_4'[0]=0$}
	\drawedge[ATnb=0,AHnb=1,ELside=l,linewidth=.5,ELpos=70](gb2,gb3){\small $x_4'[5]=1$}
	\drawedge[ATnb=0,AHnb=1,ELside=r](gb2,ga){\small 0}	
	\put(115,60){\begin{tabular}{rl}
	$x_4'$&$= x_4 ~AND~ 1 1 1 1 1 0 0 0 $ \\
	&$=1 1 {\bf 1 0} 0 0 0 {\bf 0}$
	\end{tabular}
	}
	\put(120,0){$\rk{x_4}=\rk{b}$}

	\end{picture}		
	\setlength{\unitlength}{1mm}
	\caption{
	Computing $\rk{x_4}$ after the second search in the compressed trie.}
	\label{rkt}
	\end{center}
\end{figure*}

\subsection{Fusion Tree Characteristics}

Basically, the fusion tree is a B-tree with degree $B = (\lg n)^\frac{1}{5}$, i.e, the degree is an increasing function with respect to the number of elements. Figure~\ref{fig_arvore_fusao} has an example.


\begin{figure}[htb!]
	\begin{center}
		\begin{picture}(80,50)(0,0)

	\node[Nframe=n,Nadjust=wh,Nmr=1](R)(35,48){Raiz} 
	\node[Nadjust=wh,Nmr=1](n1)(40,40){$s_1~|~s_2~|~\ldots|~s_{(\lg n)^\frac{1}{5}}$} 
	\drawedge[ATnb=0,AHnb=1](R,n1){} 
	\node[Nadjust=wh,Nmr=1](n2)(20,30){$s_1~|~s_2~|~\ldots|~s_{(\lg n)^\frac{1}{5}}$} 
	\node[Nadjust=wh,Nmr=1](n3)(60,30){$s_1~|~s_2~|~\ldots|~s_{(\lg n)^\frac{1}{5}}$} 
	\node[Nframe=n,Nadjust=wh,Nmr=1](x)(40,30){\ldots} 
	\drawedge[ATnb=0,AHnb=1](n1,n2){} 
	\drawedge[ATnb=0,AHnb=1](n1,n3){} 
	\node[Nadjust=wh,Nmr=1](n4)(15,15){$s_1~|~\ldots|~s_{(\lg n)^\frac{1}{5}}$} 
	\node[Nframe=n,Nadjust=wh,Nmr=1](n6)(33,15){$\ldots$} 
	\drawedge[ATnb=0,AHnb=1](n2,n4){}

	\imark[iangle=260,ATnb=0,AHnb=0](n2)
	\imark[iangle=270,ATnb=0,AHnb=0](n2)
	\imark[iangle=280,ATnb=0,AHnb=0](n2)
	\imark[iangle=290,ATnb=0,AHnb=0](n2)
	\imark[iangle=296,ATnb=0,AHnb=0](n2)

	\imark[iangle=245,ATnb=0,AHnb=0](n3)
	\imark[iangle=250,ATnb=0,AHnb=0](n3)
	\imark[iangle=260,ATnb=0,AHnb=0](n3)
	\imark[iangle=270,ATnb=0,AHnb=0](n3)
	\imark[iangle=280,ATnb=0,AHnb=0](n3)
	
	\node[Nframe=n,Nadjust=wh,Nmr=1](n7)(47,15){\ldots} 

	\node[Nadjust=wh,Nmr=1](n9)(65,15){$s_1~|~\ldots|~s_{(\lg n)^\frac{1}{5}}$} 
	\drawedge[ATnb=0,AHnb=1](n3,n9){}  
		\imark[iangle=245,ATnb=0,AHnb=0](n3)
	
		\imark[iangle=245,ATnb=0,AHnb=0](n3)

	\imark[iangle=255,ATnb=0,AHnb=0](n4)
	\imark[iangle=260,ATnb=0,AHnb=0](n4)
	\imark[iangle=270,ATnb=0,AHnb=0](n4)
	\imark[iangle=280,ATnb=0,AHnb=0](n4)
	\imark[iangle=285,ATnb=0,AHnb=0](n4)

	\imark[iangle=255,ATnb=0,AHnb=0](n9)
	\imark[iangle=260,ATnb=0,AHnb=0](n9)
	\imark[iangle=270,ATnb=0,AHnb=0](n9)
	\imark[iangle=280,ATnb=0,AHnb=0](n9)
	\imark[iangle=285,ATnb=0,AHnb=0](n9)

	\node[Nw=5,Nh=5,Nmr=1](n10)(10,0){} 
	\node[Nframe=n,Nw=5,Nh=5,Nmr=1](x)(15,0){\ldots} 
	\node[Nw=5,Nh=5,Nmr=1](n11)(20,0){}
	\drawedge[ATnb=0,AHnb=1](n4,n10){}
	\drawedge[ATnb=0,AHnb=1](n4,n11){} 

	\node[Nw=5,Nh=5,Nmr=1](n16)(60,0){} 
	\node[Nframe=n,Nw=5,Nh=5,Nmr=1](x)(65,0){\ldots} 
	\node[Nw=5,Nh=5,Nmr=1](n17)(70,0){} 

	\drawedge[ATnb=0,AHnb=1](n9,n16){}
	\drawedge[ATnb=0,AHnb=1](n9,n17){} 

	\end{picture}	

	\caption{Structure of a complete fusion tree.}
	\label{fig_arvore_fusao}
	\end{center}
\end{figure}

As the height of a B-tree is proportional to $\log_B n $ and the fusion tree has $B = (\lg n)^\frac{1}{5}$ so the height $h$ has complexity:


\[ \log_B n = \frac{\lg n}{\lg B} = \frac{\lg n}{\lg (\lg n)^{1/5}} = \cdot \frac{\lg n}{\lg \frac{1}{5}\lg n} = O(\frac{\lg n}{\lg \lg n}). \]


The time to search the correct child to continue the search in a B-tree node is $O(B)$ using linear search. As the search occurs in each tree level, the overall time is  $O(B \log_B n)$ to perform a search. In a fusion tree, the child is found in 
$O(1)$ in a node and in $O(\log_B n)$ in the tree. As $B =(\lg n)^\frac{1}{5}$, the search complexity time is


\[  \log_B n = O(\dfrac{\lg n}{\lg \lg n}).\]




As previously discussed, some irrelevant bits can be discarded in the sort process. A special structure name sketch is created to save only the relevant ones:

\begin{definicao}{\sk{s}:}
The sketch of a word $s$ consists in discarding all irrelevant bits, keeping the relevant ones. Sketch operations preserve the words order, i.e., $s_i < s_j$ if and only if  $\sk{s_i} < \sk{s_j}$. 
%
\end{definicao}

Figure~\ref{fig_pesquisaTrie} has the elements $a$, $b$, $c$ and $d$ sketches. They are $011$, $100$, $101$ e $110$ respectively. The sketches order doesn't change with respect to the original numbers.


The fusion tree central idea is concerned with how it stores the key in each node. Each node contains $t$ keys, for $t < B-1 = O(w^{1/5})$. As stated in Lemma~\ref{nbits}, a trie with $B-1$ keys has at most $B-2$ relevant bits. A node contains $B-1$ sketches each with $B-2$ relevant bits. Thus, the overall sketch bits in a node are
$$ (B-1)\cdot (B-2) \leq w^\frac{1}{5} \cdot w^\frac{1}{5} = O(w^\frac{2}{5}) = o(w).$$

The sum of sketches bits in a node fits in only one memory word. Thus, each fusion tree node has one word that keeps one sketch for each key plus some bits as defined above:


\begin{definicao}{Sketch Node}
The sketch node is a node that contains all keys sketches. Such sketches can be stored in only one word. Additionally, there is a separator bit between the sketches whose value is 1. The sketch node will be the concatenation of each key sketch:
$$w_{node} = 1\sk{s_1}1\sk{s_2} ... 1\sk{s_t}.$$ 
Furthermore, sketches are concatenated in nondecreasing order.
%
%
\end{definicao}

The next subsection will show how to compare a key x with all keys in a node in constant times, based in~\cite{lec2011}.

\subsection{Multiple comparisons in constant time}

%


Consider a fusion tree node with elements  $S=(s_1,\ldots,s_t)$. Suppose the relevant bits with respect to $S$ are $(i_1,\ldots,i_{t'})$ with $t'<t$. To compare a search key $x$ with all node keys, first $\sk{x}$ is computed. 

To extract the first relevant bit $i_1$ and store it in the first position of $\sk{x}$, it is computed a bitwise AND between $x$ and a mask with value 1 only in the bit $i_1$. Once the mask is applied, the bit must be moved to sketch vector position $0$. Such movement of delta bits is obtained by a multiplication by $2^{delta}$.

%
%
%

To obtain all relevant bits of $x$ in the initial position of $\sk{x}$, first a bitwise AND is performed between $x$ and a mask with 1 only in the relevant bits $(i_1,\ldots,i_{t'})$. Such mask will be constructed with the compressed trie and is available when the $x$ search is computed. After this, all relevant bits must be moved to the initial position of $\sk{x}$  in $O(1)$ as shown in the figure above.

%
%
%

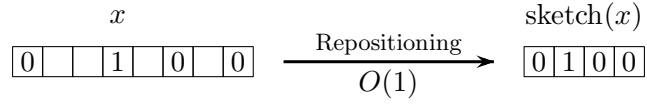
\centerline{
\begin{picture}(110,20)(0,0)
\node[Nmr=0,Nw=4,Nh=4,linecolor=white](B)(24,16){$x$}
\node[Nmr=0,Nw=4,Nh=4,linecolor=white](B)(86,16){$\sk{x}$}
\node[Nmr=0,Nw=4,Nh=4](B)(12,10){0}
\node[Nmr=0,Nw=4,Nh=4](B)(16,10){}
\node[Nmr=0,Nw=4,Nh=4](B)(20,10){}
\node[Nmr=0,Nw=4,Nh=4](B)(24,10){1}
\node[Nmr=0,Nw=4,Nh=4](B)(28,10){}
\node[Nmr=0,Nw=4,Nh=4](B)(32,10){0}
\node[Nmr=0,Nw=4,Nh=4](B)(36,10){}
\node[Nmr=0,Nw=4,Nh=4](B)(40,10){0}
\node[Nframe=n,Nadjust=wh](Y)(75,10){} 	
\node[Nframe=n,Nadjust=wh](X)(45,10){} 	
\drawedge[ATnb=0,AHnb=1,linewidth=.4,AHLength=2.5](X,Y){\footnotesize Repositioning}
\drawedge[ELside=r](X,Y){$O(1)$} 			 	
\node[Nmr=0,Nw=4,Nh=4](B)(80,10){0}
\node[Nmr=0,Nw=4,Nh=4](B)(84,10){1}
\node[Nmr=0,Nw=4,Nh=4](B)(88,10){0}
\node[Nmr=0,Nw=4,Nh=4](B)(92,10){0}
\end{picture}
}

When an arbitrary number $x$ is multiplied by a predefined constant, it is possible to reposition the bits of $x$. The problem of repositioning the relevant bits of $x$ to the initial position of $\sk{x}$ in $O(1)$ is a nontrivial task. The work~\cite{lec2012} discuss the existence of predefined constants to reposition the relevant bits of $x$. The result is imperfect because some additional zeros bits are added to the $\sk{x}$. Such additional zero bits do not change the algorithm behavior. The $\sk{x}$ computation is not covered by this work.


Once $\sk{x}$ is computed, its value is concatenated $t$ times in the following way:
 $$w_x = 0~\sk{x}~0~\sk{x} ~...~ 0~\sk{x}.$$

Suppose that $\sk{x}$ has 6 bits, so:
\begin{equation*} 
\begin{aligned}
w_x &= \sk{x} + \sk{x}\cdot 2^7 + \sk{x} \cdot 2^{14} + ~... \\
& = \sk{x}\cdot(...10000010000001).\\
\end{aligned}
\end{equation*}

Thus, $w_x$ is computed from $\sk{x}$ with only one multiplication.

\begin{fato}
When subtracting $1\sk{s_i}-0\sk{x}$, the result starts with 1 if and only if $\sk{x}\leq \sk{s_i}$.
\end{fato}

Let $\sk{x} = 1111$ and $\sk{s_i} = 0000$, thus $1\sk{s_i} - 0\sk{x} ~ = ~ 10000 - 01111 = \mathbf{0}0001$. As the subtraction result starts with zero, then $\sk{x}>\sk{s_i}$.


Suppose $\sk{x}=0000$ and $\sk{s_i}=00001$. Thus $1\sk{s_i} - 0\sk{x} = 10001-00000 = \mathbf{1}0001$. As the subtraction result starts with 1, $ \sk{x}\leq \sk{s_i}  $.
%

To compare $x$ with all words in $S$ in $O(1)$, a subtraction between $1\sk{s} - 0\sk{x}$ for all  $s\in S$ is performed in one operation. That means many comparisons with only one operation.
The calculus is 
$$w_{res} = w_{node}-w_x.$$

The first bit of each block will indicate if $\sk{x}$ is lesser than or equal or greater than $\sk{s_i}$. As the sketches are sorted in a $w_{node}$, the first block that starts with 1 must be found. Suppose that the number of bits of a block $0\sk{x}$ is $r$. To remove all bits except the first bit of each block, a bitwise AND is performed between $w_{res}$ and a mask with value 1 in the positions $r$, $2r$, $3r$ and so on. Let $w'_{res}$ the result of such bitwise AND. The next step consists in finding the most significant bit that values 1. Such operation is equivalent to calculate $\lfloor\lg (w'_{res})\rfloor$ and must be performed in O(1). Such problem is found in the literature~\cite{hackers}.

%

The element $s=\trs{x}$  can be computed from the position of the first 1 in a $w'_{res}$. Following the steps of previous section, the rank value can be computed in $O(1)$. Thus the correct child to continue the search in a fusion tree is computed in $O(1)$.

The following example will provide a step-by-step execution of $\trs{x}$ with the same values used in Figure~\ref{fig_pesquisaTrie}:
%
$$S=(a,b,c,d)=(11011111, 11100000, 11100001, 11111110)=(223,224,225,245)$$

Consider $x = 11100111=231$. The sketches will be 
$$ \sk{a} = 011;~~~ \sk{b} = 100; $$
$$\sk{c} = 101;~~~ \sk{d} = 110. $$

The sketch node will be
$$ w_{node} = 1~011~1~100~1~101~1~110=48350,$$
and the $\sk{x}$ will be $101$. The word to subtract from  $ w_{node}$ will be:
$$ w_q = 0~101~0~101~0~101~0~101 =21845.$$

Performing the subtraction and applying a bitwise AND it will result in:

\begin{equation}
\begin{aligned}
w_{res}& = (w_{node} - w_q) ~AND~ 1~000~1~000~1~000~1~000, \\
& =(48350-21845)~AND~ 1~000~1~000~1~000~1~000, \\
& =26505~AND~ 1~000~1~000~1~000~1~000,\\
& = 0~110~0~111~1~000~1~001~~AND~1~000~1~000~1~000~1~000,  \\
& = 0\_\_\_0\_\_\_{\bf 1}\_\_\_1\_\_\_  = 136.
\end{aligned}
\end{equation}

The first bit that values 1 is $\lfloor\lg (136)\rfloor=b_7$. The bits are zero indexed, thus there are 8 bits from the beginning to the first 1. Such value divided by the block size 4 results in 2, which 
is the penultimate element in the sequence $(s_0,s_1,s_2,s_3)$. Thus,$\trs{x}=c$ because $S=(a,b,{\bf c},d)$.

%

\subsection{
Sorting in  $o(n \lg n)$
}

This work detailed how to search a $w$-bit word in  $O(\frac{\lg n}{\lg \lg n})$  in a fusion tree data structure. It also describes how to sort $n$ elements using B-tree.  All elements inserted in a fusion tree result in a sorted set of elements.
The paper~\cite{dynamic} shows how to transform a static fusion tree in a dynamic one. A dynamic fusion tree is optimized to update keys in $O(\frac{\lg n}{\lg \lg n} + \lg(\lg(n)))$ by update. The resulting sort complexity is
\[  n \left(\log_B n + \frac{\lg n}{\lg \lg n}+\lg\lg n \right) = \]
\[  O\left(n \frac{\lg n}{\lg \lg n}\right).  \]

\section{Conclusion}
\label{cap:conclusao}

This work aimed to describe the fusion tree data structured and the $O(\frac{n\lg n}{\lg \lg n})$ sorting algorithm. Step by step examples is prepared for didactic purposes. Very few materials are available related to this relevant issue. The challenge was to understand many theorems and non trivial concepts and prepare a material to a wide community.

	This work let some open questions as \textit{(i)} how to discover the first bit 1 in $w$-bit word in $O(1)$; \textit{(ii)} how to compute $\sk{x}$ in $O(1)$ and \textit{(iii)} how to create dynamic fusion tree optimized to update keys. Anyway, this work successfully completes the task of detailing the fusion tree data structure, responsible for the first $o(n \lg n)$ sorting algorithm and a basis for many other subsequent algorithms.
	
Such work also reveals some pitfalls in the use of lower bounds. For instance, if a generic problem needs at least $f(n)$ operations, the real lower bound is $\Omega(f(n)/\lg n)$ because the widely accepted computational models are able to process $\lg n$ bit in $O(1)$.

An opportune future work would be to implement the fusion tree sorting algorithm and compare it with traditional algorithms. Another relevant aspect is the possibility of multiple operations in $O(1)$ and the removal of irrelevant bits. Such possibilities present theoretical and practical consequences. In the theoretical field, the question is which problems could have their complexity decreased with multiple operations in $O(1)$. In applied computing, the use of multiple operations inside a single word and the removal of irrelevant bits can accelerate traditional algorithms.

\bibliography{bibl}
\bibliographystyle{plain}


\end{document}